\documentclass[amsmath, amssymb, preprintnumbers, showpacs, showkeys,aps,prb,twocolumn]{revtex4-1}
\usepackage{graphicx}
\usepackage{float}
\usepackage{color}
\usepackage{braket}
\usepackage{ulem}   
\usepackage{slashed}
\usepackage{mathtools}
\usepackage{bbm}    
\newcommand{\bs}[1]{{\boldsymbol{#1}}}

\newcommand{\wt}{WTe$_2$}

\newcommand{\bcx}{\bar{C}_{2x}}

		\newcommand{\e}[1]{\begin{align}{#1}\end{align}}	
		\newcommand{\f}[2]{\frac{#1}{#2}}

		\newcommand{\la}[1]{\label{#1}}
		\newcommand{\q}[1]{Eq.\ (\ref{#1})}
		\newcommand{\s}[1]{Sec.\ \ref{#1}}
		\newcommand{\fig}[1]{Fig.\ \ref{#1}}		

\newcommand\myspace{\;\;\;\;}

\newcommand{\be}{\boldsymbol{e}}

\newcommand{\bk}{\boldsymbol{k}}

\newcommand{\br}{\boldsymbol{r}}

\newcommand{\bA}{\boldsymbol{A}}
\newcommand{\bB}{\boldsymbol{B}}

\newcommand{\bG}{\boldsymbol{G}}

\newcommand{\bR}{\boldsymbol{R}}

\newcommand{\bdelta}{\boldsymbol{\delta}}

\newcommand{\bmx}{\bar{M}_x}

\newcommand{\hatgdel}{\pdg{\hat{g}}_{\sma{\bdelta}}}

\newcommand{\ins}[1]{\;\;\;\;\text{#1}\;\;\;\;}

\newcommand{\calh}{{\cal H}}
\newcommand{\cali}{{\cal I}}

\newcommand{\calp}{{\cal P}}

\newcommand{\noi}[1]{\noindent (#1)}

\newcommand{\mo}{\text-1}

\newcommand{\pdg}[1]{{#1}^{\phantom{\dagger}}}

\newcommand{\lin}{\notag \\}

\newcommand{\eq}{=&\;}

\newcommand{\low}{L$\ddot{\text{o}}$wdin\;}

\newcommand{\W}{{\cal W}}

\newcommand{\R}{\mathbb{R}}

\newcommand{\dg}[1]{#1^{\scriptstyle{\dagger}}}

\newcommand{\sma}[1]{\scriptscriptstyle{#1}}

\newcommand{\Z}{\mathbb{Z}}

\newcommand{\qed}{\nobreak \ifvmode \relax \else
      \ifdim\lastskip<1.5em \hskip-\lastskip
      \hskip1.5em plus0em minus0.5em \fi \nobreak
      \vrule height0.75em width0.5em depth0.25em\fi}

\begin{document}
\preprint{}

\title{Topological metals from band inversion}
\author{Lukas Muechler$^1$}
\author{A. Alexandradinata$^{2,3}$}
\author{Titus Neupert$^4$}
\author{Roberto Car$^{1,3}$}
\affiliation{
$^1$Department of Chemistry, Princeton University, Princeton, New Jersey 08544, USA\\
$^2$Department of Physics, Yale University, New Haven, Connecticut 06520, USA\\
$^3$Department of Physics, Princeton University, Princeton, New Jersey 08544, USA\\
$^4$Princeton Center for Theoretical Science, Princeton University, Princeton, New Jersey 08544, USA}

\date{\today}

\begin{abstract}
We expand the phase diagram of two-dimensional, nonsymmorphic crystals at integer fillings that do \emph{not} guarantee gaplessness. In addition to the trivial, gapped phase that is expected, we find that band inversion leads to an unexplored class of topological, gapless phases. These topological phases are materialized in in the monolayers of MTe$_2$ (M $=$ W, Mo) if spin-orbit coupling is neglected. 
Furthermore, we characterize the Dirac band touching of these topological semimetals by the Wilson loop of the non-Abelian Berry gauge field. An additional feature of the Dirac cone in monolayer MTe$_2$ is that it tilts over in a Lifshitz transition to produce electron and hole pockets, a type-II Dirac cone. These pockets, together with the pseudospin structure of the Dirac electrons, suggest a unified, topological explanation for the recently-reported, non-saturating magnetoresistance in WTe$_2$, as well as its circular dichroism in photoemission.
We complement our analysis and first-principle bandstructure calculations with an \textit{ab-initio}-derived tight-binding model for the WTe$_2$ monolayer.

\end{abstract}
\maketitle


\section{Introduction}

In the Landau-Ginzburg paradigm,\cite{Ginzburg_Landau} different phases of matter are distinguished by their symmetry. A recent major advance led to the recognition that insulators with the same symmetries and particle numbers can be topologically distinct.\cite{zak1989,Haldane1988,kane2005B} That is, for the same integer electron filling and symmetry class, one may have either trivial or topological insulators. The latter have unusual electronic properties that originate\cite{Cohomological} from the Berry phase\cite{berry1984,zak1989} of electronic wavefunctions. This scenario is substantially modified for a broad class of crystals having nonsymmorphic symmetries, namely, the spatial symmetries that unavoidably translate the spatial origin by a fractional lattice vector.\cite{Lax} Nonsymmorphic symmetries guarantee that at certain integer fillings, the phase of matter must always be gapless. This robust and unavoidable semimetallicity originates from the nontrivial connectivity\cite{connectivityMichelZak,Young} of elementary energy bands.\cite{elementaryenergybands} 

\begin{figure}[t]
 \centering
 \includegraphics[width=1\columnwidth]{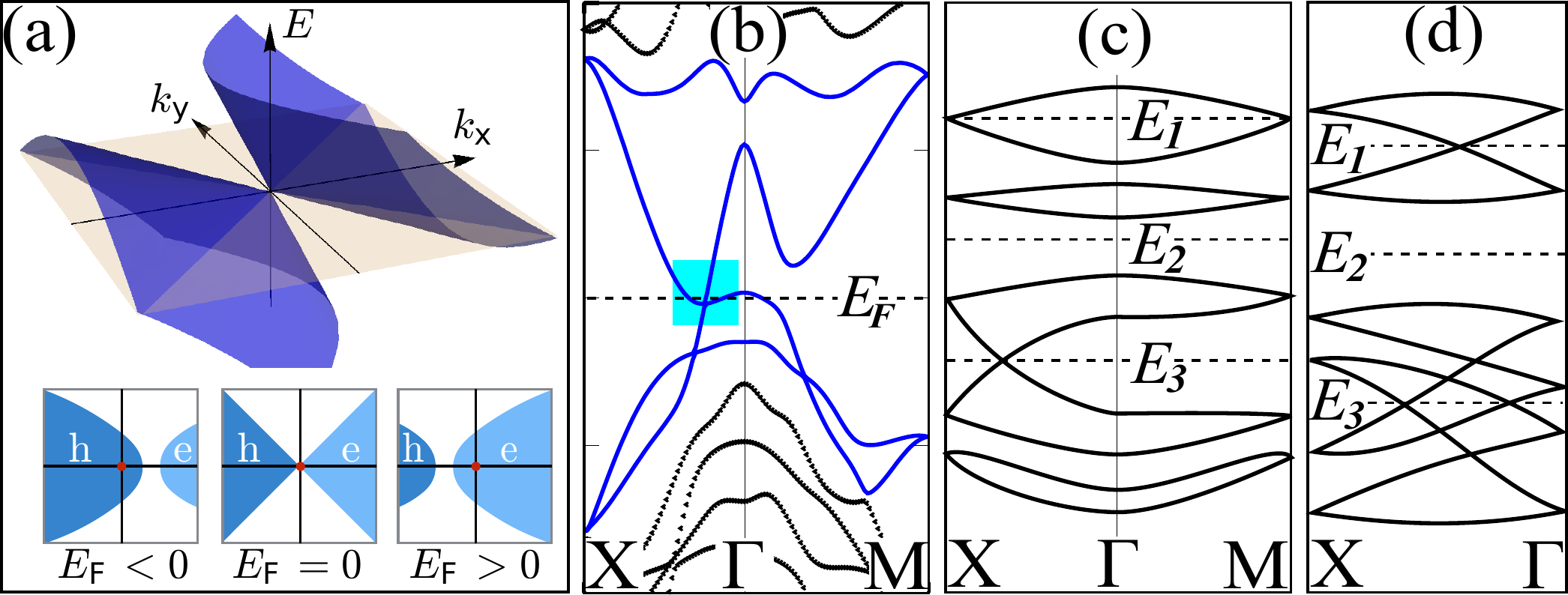}
 \caption{(a) A Dirac fermion that tilts over. (b) Bandstructure
of a \wt~monolayer without spin-orbit coupling; the tilted
Dirac fermion is highlighted by a blue square. (c) Bandstructure of a nonsymmorphic, spin-orbit-free crystal at various fillings. (d)
Bandstructure of a nonsymmorphic, spin-orbit-coupled crystal
at various fillings. For both (c-d), $E_1$, $E_2$, $E_3$ are Fermi energies that respectively exemplify (1) a semimetal whose metallicity originates from its filling, (2) a generic insulator, and
(3) a topological semimetal whose metallicity originates from
a band inversion.
}
\label{fig_1}
\end{figure}

Our work explores the distinct phases of nonsymmorphic matter, but for integer fillings that do not guarantee gaplessness, as illustrated in Fig.~\ref{fig_1}(c-d). We find that a band inversion separates a trivial, gapped phase from a topological, gapless phase. The latter semimetal is concretely exemplified by MTe$_2$ (M = W, Mo) monolayers, as we substantiate with ab-initio calculations and tight-binding models. 
We characterize this metal by a topological invariant based on the Wilson loops of the non-Abelian Berry gauge field, which contrasts with previous Abelian Berry-phase characterizations of topological semimetals.\cite{wan2010} Wilson loops not only fundamentally characterize band topology through holonomy,\cite{AA2014} they have recently emerged as an efficient method to diagnose topological \emph{insulators} from first-principles calculations.\cite{AA2014,yu2011,alexey2011,Maryam2014,berryphaseTCI} We propose here that the Wilson loop is also a powerful tool to identify and characterize topological \emph{metals}.

The low-energy excitations of MTe$_2$ monolayers are described by Dirac fermions that are topologically distinct from the rotationally-symmetric Dirac fermions in graphene. Precisely, the Dirac cone disperses so anisotropically that it `tilts over' in a Lifshitz transition, i.e., part of the upper Dirac cone dips below the nodal energy ($\epsilon = 0$) as illustrated in Fig.~\ref{fig_1}(a), resulting in a discontinuous change in the band contours at the nodal energy. We refer to it as a type-II Dirac cone, in analogy with a notion recently introduced for Weyl semimetals~\cite{WTe2Weyl} that is, incidentally, materialized by 3D MTe$_2$\cite{WTe2Weyl,MoTe1,MoTe2,MoTe3}. Instead of a point-like Fermi surface where the Fermi energy lies at a (potentially tilted) Dirac node\cite{ttf1,genchiral,goerbig,tiltedtransport,Bergholtz1,FeAsDirac,BiTeI,WeylIISuper}, a type-II Dirac cone is characterized by electron- and hole-like Fermi surfaces that touch at the Dirac node~\cite{nagaosatilt}[see Fig.~\ref{fig_1}(a)]. This novel scenario promises an abundance of unexplored experimental possibilities.\cite{Type2breakdown} Our theory and tight-binding models should serve as important resources for ongoing experimental efforts\cite{Wte2monoraman,Wte2monocond} focused on the synthesis and the study of MTe$_2$ monolayers.

The type-II Dirac fermions in spin-orbit-free WTe$_2$ monolayers provide a unifying explanation for many phenomena in spin-orbit-coupled monolayers, bilayers and 3D layered \wt. The latter are materials that have been fabricated and are under intense experimental scrutiny because of their giant, non-saturating transverse magnetoresistance (MR) of $13 \times 10^6 $~\%  at 0.53~K and 60~T, with new crystals achieving  $1.7 \times 10^6 $ ~\% at 2~K and 9~T.\cite{Wte2nature,EPLWTe2} Moreover, the angle-resolved photoemission (ARPES) of \wt~exhibits circular dichroism (CD),\cite{Wte2CD} a phenomenon which typifies Dirac semimetals such as graphene.\cite{CDgraphene1,CDgraphene2} 
To summarize our results:

\noi{i} When spin-orbit coupling is introduced, the degeneracy of the Dirac points is lifted and the disconnected bands are then topologically non-trivial \cite{qian2014quantum} in the time-reversal-symmetric $\mathbb{Z}_2$ classification.\cite{kane2005B,QSHE_Rahul,QSHE_bernevig} In this work, we propose a criterion on the  spin-orbit-free semimetal which is equivalently expressed by the number of Dirac fermions or the eigenvalues of the nonsymmorphic symmetry. If this criterion is satisfied, as is the case for \wt, spin-orbit coupling induces $\Z_2$ topological order. We remark that spin-orbit-coupled \wt~remains semimetallic due to the persistence of its electron and hole pockets, which, again, originate from the tilted Dirac fermion. 

\noi{ii} In bilayer WTe$_2$, the coupling between the two stacked monolayers breaks the nonsymmorphic symmetry that protects the Dirac fermions. The low-energy theory is then described by tilted Dirac fermions with small masses; the two-component wavefunction at each Fermi circle forms a pseudospin that rotates around the Dirac node, where Berry curvature\cite{berry1984} is concentrated.

\noi{iii} The electron and hole pockets, Berry curvature and rotating pseudospin are retained in 3D WTe$_2$, which comprises weakly-coupled bilayers. We propose that the high mobilities in transport experiments should be attributed to suppressed backscattering due to the rotating pseudospin, while the observed circular dichroism should be correlated with the Berry phase of the Dirac cones.\cite{CDgraphene1,CDgraphene2} 

This work is organized as follows: after a preliminary description of the nonsymmorphic symmetries of MTe$_2$ in \s{sec:spacegroup}, we introduce the theory of band-inverted topological semimetals in \s{sec:theory}. In \s{sec: mono} we exemplify our theory with monolayer \wt, for which we present a tight-binding model and introduce the notion of type-II Dirac cones that tilt over. We then extend our discussion to bilayer  \wt~in \s{sec: bi}, with focus on its dichroism. In \s{sec:discussmr}, we summarize our results and further relate them to the  magnetoresistance measurements in 3D WTe$_2$. Details on the derivation of the topological invariant, the role of spin-orbit coupling, the tight-binding model, and the CD calculation are collected in App.~\ref{app:wilson}, ~\ref{app:qshi}, \ref{app: tight-binding}, and \ref{app: CD}.

\section{nonsymmorphic space groups and relevant crystal structures} \label{sec:spacegroup}

In crystals, a basic geometric property that distinguishes spatial symmetries concerns how they transform the spatial origin: rotations, inversions and reflections preserve the origin, while screw rotations and glide reflections unavoidably translate the origin by a rational fraction of the lattice period.\cite{Lax} If no origin exists that is simultaneously preserved, modulo integer lattice translations, by all the symmetries in a space group, this space group is called nonsymmorphic. In \s{sec:crystalmono} we exemplify a nonsymmorphic space group with the crystal structure of MX$_2$ monolayers , which applies to WTe$_2$, MoTe$_2$, and ZrI$_2$. 
In contrast, the MX$_2$ bilayer is characterized by a symmorphic space group, as we explain in \s{sec:crystalbi}.  

\subsection{Crystal structure of the MX$_2$ monolayer} \la{sec:crystalmono}

The $M$ atoms form zigzag chains along $\bs{e}_x$, and are coordinated by $X$ atoms that form distorted edge-sharing octahedra; here $\bs{e}_x,\bs{e}_y,\bs{e}_z$ are basis vectors in a Cartesian coordinate system, with $\bs{e}_z$ orthogonal to the monolayer, and $\bs{e}_x,\bs{e}_y$ the generators of the Bravais lattice of the monolayer; the lengths of $\bs{e}_x,\bs{e}_y$ correspond to lattice constants that we denote respectively by $a,b$. We label a unit cell in the monolayer by $\bs{R} = h \bs{e}_x + k \bs{e}_y $ with $h,k \in \Z$.

\begin{figure}[t]
\centering
\includegraphics[width=\columnwidth]{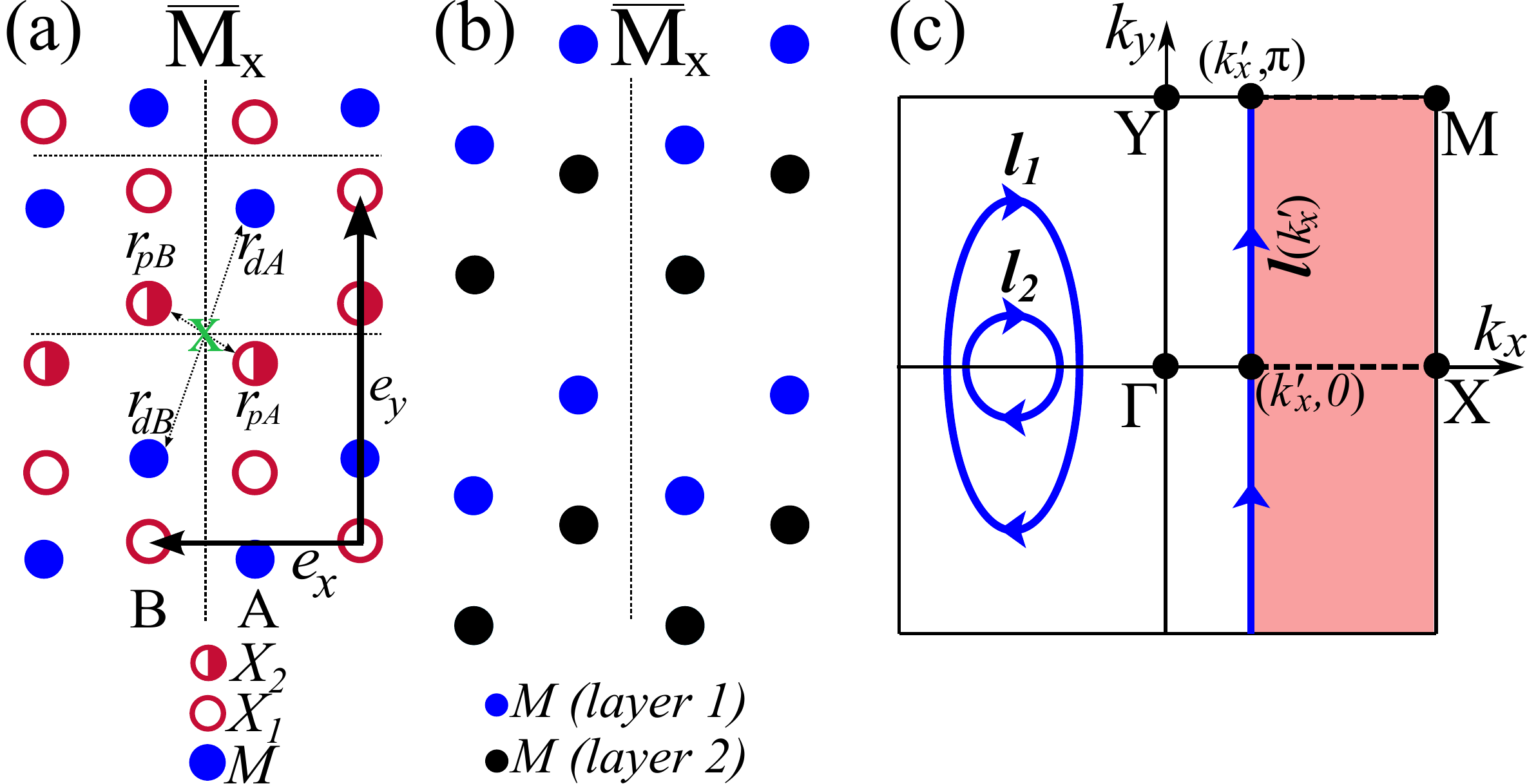}
\caption{
(a) Crystal structure of a MX$_2$ monolayer. The horizontal lines are invariant under the screw $\bcx$; the vertical line is invariant under $\bmx$, and the green cross indicates the center of inversion and also our choice of spatial origin. Within each unit cell (encircled by the rectangle), we divide the four $X$ atoms into two pairs marked $X$-1 and $X$-2; The $X$-2 atoms are further divided into two sublattices labelled by A and B. $\br_{\ell,s}$, with $\ell \in \{d,p\}$ and $s\in\{A,B\}$, are vectors connecting the origin to the centers of the Wannier functions that we introduce in \s{sec:tbspinless}. 
(b) Crystal structure of a MX$_2$ bilayer, where only the $M$ atoms are shown. (c) Brillouin zone of both the monolayer and the bilayer. In \s{sec:theory}, we characterize the monolayer by screw-symmetric Wilson loops indicated in blue: $l_1$ and $l_2$ are contractible, while $l(k_x')$ is not.}
\label{fig:wilsonloops}
\end{figure}

The group of a MX$_2$ monolayer is generated by (i) time reversal ($T$), (ii) lattice translations $t(\bs{e}_x)$ and $t(\bs{e}_y)$, where $t(\br)$ indicates a translation by the vector $\br \in \R^3$, as well as (iii) a reflection $\bar{M}_x \equiv t(\bs{e}_x/2)M_x$, which is a product of a reflection $M_x$, acting as $M_x{:}(x,y,z)\rightarrow (-x,y,z)$, and a translation $t(\bs{e}_x/2)$ by half a lattice vector,  and (iv) a screw rotation $\bar{C}_{2x} = t(\bs{e}_x/2)C_{2x}$, which is the product of a two-fold rotation $C_{2x}$, acting as $C_{2x}{:}(x,y,z) \rightarrow (x,-y,-z)$, and the same fractional translation. The product of the last two generators is the spatial inversion $\bmx \bcx = \cali$ that sends $\bs{r}\rightarrow-\bs{r}$; we choose the inversion center, indicated by a green cross in Fig.\ \ref{fig:wilsonloops}(a), as our spatial origin. The atomic positions of the freestanding monolayers that we considered were extracted from monolayers within 3D crystals, whose parameters are known from experiments.\cite{WTe2struc,ZrI2struc,MoTe2Weyl2} 

\subsection{Crystal structure of the MX$_2$ bilayer} 
\la{sec:crystalbi}


Let the position of each atom in a MX$_2$ monolayer be parametrized by $(x,y,z)$ relative to our chosen spatial origin (green cross Fig.~\ref{fig:wilsonloops}(a)). There is a  corresponding identical atom positioned at $(-x+\frac{1}{2}+a,-y+b,z+c)$ in the second layer of the bilayer, where $\{a,b,c\}$ are material-specific parameters.
This stacking spoils both inversion and screw symmetries of the monolayer, but retains the mirror symmetry $\bar{M}_x$, as we illustrate in Fig.~\ref{fig:wilsonloops}(b). Our stacking is identical to that of the bilayer within 3D MX$_2$,  whose experimentally known atomic positions \cite{WTe2struc} we use throughout this paper.

\section{Theory of band-inverted topological semimetals} \la{sec:theory}

In \s{sec:comparingsemimetals}, we briefly review nonsymmorphic semimetals which are semimetallic only due to their filling. These semimetals are distinguished from semimetals which originate from band inversion, as we briefly describe in \s{sec:comparingsemimetals}, and then more carefully elaborate in \s{sec:bandinversion}. These band inverted semimetals admit a topological classification that we describe in \s{sec:top}. When a gap is induced by spin-orbit coupling, certain band inverted semimetals turn into topological insulators, as we substantiate in \s{sec:qshi}.

\subsection{Comparing topological and filling-enforced semimetals} \la{sec:comparingsemimetals}

We have introduced two types of nonsymmorphic semimetals: 
(i) \emph{filling-enforced semimetals}, which are guaranteed to be semimetallic at certain fillings determined by the space group;\cite{connectivityMichelZak}
(ii) \emph{topological semimetals}, which are not guaranteed in the sense of (i), but are semimetallic due to a topological band inversion that we will describe. To exemplify (i) and (ii), we offer two examples from a group generated by $\{T,t(\bs{e}_x),t(\bs{e}_y), \bcx\}$, which has one less generator ($\bmx$) than the group of WTe$_2$; we have defined this smaller group to emphasize the relevant symmetries, as well as their wider applicability to other materials. 

In this Section, we consider electronic systems without spin-orbit coupling. In our definition of filling ($f$), we count a spin-degenerate band as a single band, and each spin species transforms in an integer-spin representation\cite{Lax,tinkhambook} of the space-group symmetries described in Sec.\ \ref{sec:spacegroup}, e.g., $\bcx^2=t(\bs{e}_x)$ would not include a $2\pi$ rotation of the spin. There are two lines ($k_y=0$ and $\pi$) which are individually mapped onto themselves under the screw transformation; in short, we call them screw lines. Bands along each screw line may be labelled by the eigenvalues of $\bcx$, which fall into the two momentum-dependent branches, $\pm\mathrm{exp}(-ik_x/2)$, as follows from $\bcx^2=t(\bs{e}_x)=\mathrm{exp}(-i k_x)$ in a Bloch-wave representation. 
At inversion-invariant points $X$ [$(k_x,k_y)=(\pi,0)$] and $M$ [$(k_x,k_y)=(\pi,\pi)$] on the screw lines, 
time reversal pairs up complex-conjugate representations of $\bcx$, such that the bands are all doubly-degenerate. Each degenerate subspace is composed of an equal number of states with $\bcx$-eigenvalue $+ i$ and $-i$. In contrast, time reversal does not enhance the degeneracy at $\Gamma$ [$(k_x,k_y)=(0,0)$] and $Y$ [$(k_x,k_y)=(0,\pi)$] where the $\bcx$-eigenvalues are real. The situation is illustrated in Fig.\ \ref{fig_1}(c): bands divide minimally into pairs, such that within each pair there is at least one robust contact point (here, a crossing between orthogonal screw representations at $X$/$M$) that connects both members of the pair -- we say that bands are two-fold connected along both screw lines.\cite{Hourglass} The notion of connectivity of a submanifold (here, a screw line) generalizes\cite{Hourglass} the notion of symmetry-enforced degeneracy at an isolated wavevector; the connectivity of the entire Brillouin zone\cite{connectivityMichelZak} relates to the theory of elementary energy bands.\cite{elementaryenergybands} Due to the two-fold connectivity, any odd, single-spin filling ($f\in 2\Z+1$) is guaranteed to produce a \emph{filling-enforced} semimetal, as exemplified by the Fermi energy $E_1$ in Fig.\ \ref{fig_1}(c).

If the filling is even ($f\in 2\Z$), semimetallicity is not guaranteed by reason of filling, as exemplified by the Fermi energy $E_2$ in Fig.\ \ref{fig_1}(c). Dirac semimetallicity is nevertheless guaranteed at even filling due to independent band inversions along either screw line, e.g., a single inversion at $\Gamma$ (resp.\ Y) would nucleate a pair of time-reversal-related Dirac crossings that situate anywhere along $X\Gamma X$ (resp.\ along $MYM$), as we illustrate with the spin-orbit-free WTe$_2$-monolayer in Fig.\ \ref{fig_1}(b). The general theory of band-inverted semimetals is elaborated in the next section. We remark that a glide reflection $\bar{M}_y$, composed of a reflection $(y\rightarrow-y)$ and a half-lattice translation in $\bs{e}_x$, also satisfies $\bar{M}_y^2=t(\bs{e}_x)$, just like $\bcx$. Consequently, every result in this and the next two Sections applies also to $\bar{M}_y$, with the cosmetic substitution `screw' $\rightarrow$ `glide'.

\subsection{Band inversion and Dirac semimetallicity at even filling} \la{sec:bandinversion}

\begin{figure}[t]
 \centering
 \includegraphics[width=0.95\columnwidth]{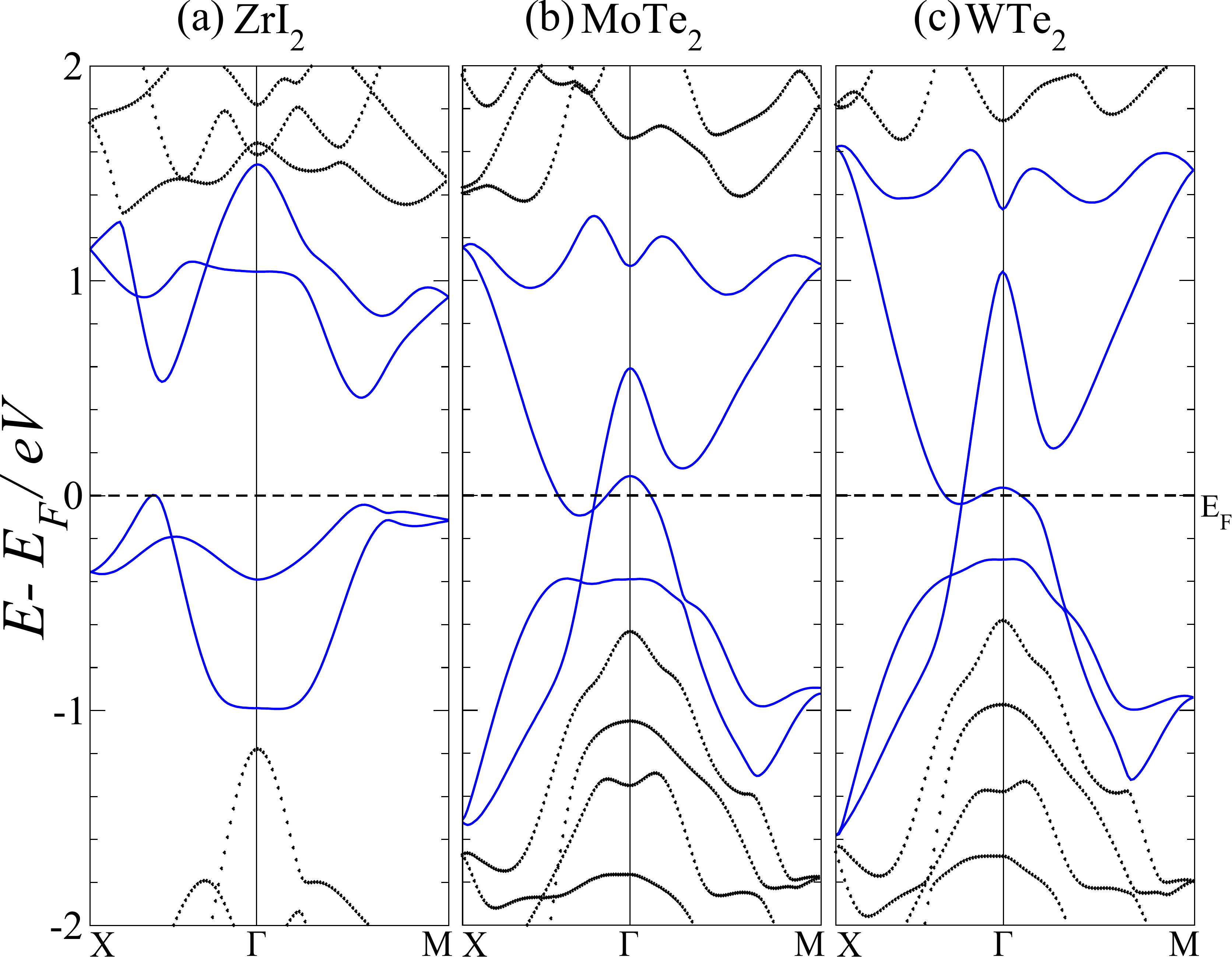}
 \caption{DFT band structures of monolayer ZrI$_2$ (a), MoTe$_2$ (b) and WTe$_2$ (c) calculated without spin orbit coupling. Bands close to the Fermi level have been plotted in blue to emphasize the different connectedness of the bands between ZrI$_2$ and MoTe$_2$/WTe$_2$.}
\label{fig_2}
\end{figure}

In this Section, we focus on even filling and quantify the relation between band inversion and Dirac semimetallicity. We say bands are inverted if the filled states, at any wavevector along a screw line, transform nontrivially under the screw rotation. To further clarify `filled states at a wavevector' for a filling $f$, we say that a state at a wavevector is filled if it belongs to the lowest set of $f$ bands at that wavevector. Filled states according to our unconventional notion of filling often coincide with actual states below the Fermi level [e.g., ZrI$_2$ in Fig.\ \ref{fig_2}(a)]; exceptions include the hole/electron pockets of \wt~and MoTe$_2$ [Fig.\ \ref{fig_2}(b-c)], which originate from the Dirac fermion tilting over, as we will elaborate in Sec.\ \ref{sec:typeII}. Our notion of filled states more naturally generalizes to bosonic systems, as we will elaborate. Furthermore, we now demonstrate how the symmetry analysis of filled states is predictive of the number of Dirac crossings, whether or not they tilt over.

\begin{figure}[t]
\centering
\includegraphics[width=7 cm]{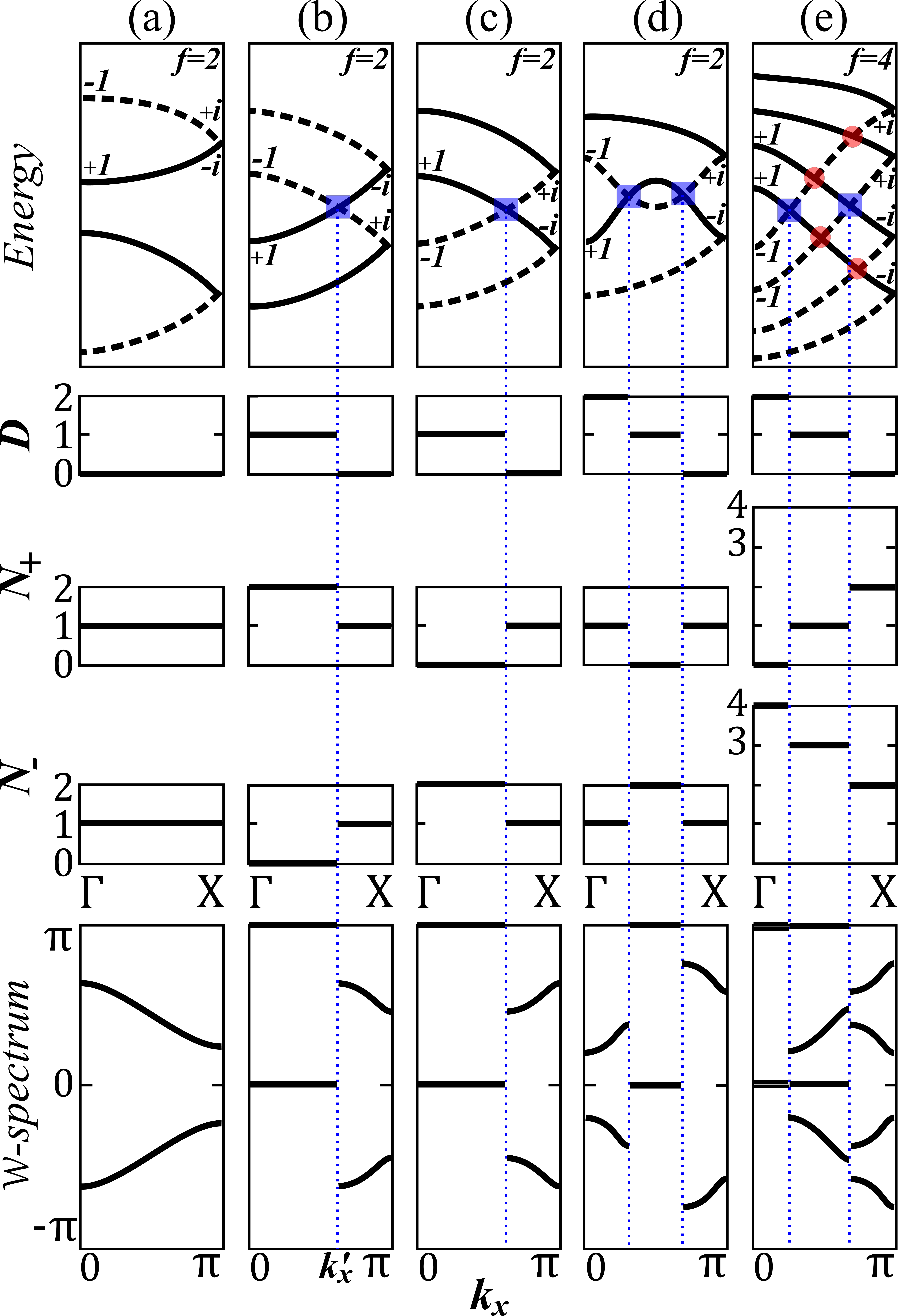}
\caption{Characterization of a generic insulator (a) and various topological semimetals (b-e).  Top row: bandstructures along the screw line $\Gamma X$. Solid (dashed) lines correspond to bands in the odd (even) screw representation, i.e., having  screw eigenvalue $+\mathrm{exp}(-ik_x/2)$ [$-\mathrm{exp}(-ik_x/2)$]. For each of (a-e), the filling ($f$) is indicated in the top-right corner, and we assume that there are no Dirac crossings along the other screw line $YM$. Blue circles denote crossings between a filled and an unfilled band, whereas red circles mark crossing between bands that are both filled (unfilled).  Second, third and fourth rows: respectively, the numbers of Dirac crossings ($D_{\sma{\bk}}$), of even, filled states ($N_{\sma{+,\bk}}$), and of odd, filled states ($N_{\sma{-,\bk}}$), for $\bk$ along $\Gamma X$. Last row: eigenvalue-phases of the $\W(k_x)$-spectrum. }\label{fig:bandinversion}
\end{figure}

Precisely, we would count, for any $\bk$ along $X\Gamma X$ (resp.\ $MYM$), the number of Dirac crossings along the screw line that connects $\bk$ to $X$ (resp.\ $M$) in the direction of increasing $k_x$, as indicated by the bottom (resp.\ top) dashed line in \fig{fig:wilsonloops}(c). Henceforth, we refer to this number as the Dirac count $(D_{\sma{\bk}})$ e.g., $D_{\sma{\Gamma}}$ is the number of crossings along half the screw line: $\Gamma X$. For illustration, we plot $D_{\sma{(k_x,0)}}$ for five case studies in Fig.\ \ref{fig:bandinversion}. To further clarify $D_{\sma{\bk}}$, we consider only the crossings between the top-most, filled band and the bottom-most, unfilled band, as highlighted by blue squares in Fig.\ \ref{fig:bandinversion}(b-e), i.e., we discard crossings that may occur between two filled bands, and also those between two unfilled bands [e.g., red circles in Fig.\ \ref{fig:bandinversion}(e)].

To characterize the symmetry representation of the filled states, let us denote by $N_{\sma{+,\bk}}$ ($N_{\sma{-,\bk}}$) the number of filled, even (odd) Bloch states at $\bk$ along either screw line, where the even (odd) representation is defined to have screw eigenvalue $+$exp$(-ik_x/2)$ [$-$exp$(-ik_x/2)$]; $N_{+,\bk}+N_{-,\bk}=f$, and where $N_{+,\bk} \neq N_{-,\bk}$, we say that bands are inverted at $\bk$, as exemplified by the interval $0 \leq k_x <  k_x'$ in \fig{fig:bandinversion}(b). $N_{\sma{\pm,\bk}}$ relate to the total Dirac count through
\e{
D_{\bk} = \f{|N_{+,\bk}-N_{-,\bk}|}{2} + 2n_{\bk};\;\; n_{\bk} \in \Z^{\sma{\geq}}. \la{genD}
} 
That is, $n_{\sma{\bk}}$ belongs to the set of nonnegative integers, and is undeducible solely from $N_{\sma{\pm,\bk}}$. To prove \q{genD} for $\bk$ along $X\Gamma X$, recall that each connected pair of bands (where, again, each pair corresponds to two eigenvalue-branches of $\bcx$ which unavoidably cross in the interpolation: ($k_x\rightarrow k_x+2\pi$) is time-reversal-degenerate at $X$, with each degenerate subspace comprising a single odd ($\bcx=+i$) state and a single even state with $\bcx=-i$. Given an even-integral filling ($f \in  2\Z$), there are then an equal number ($f/2$) of filled even and filled odd states at $X$. Since each band is smoothly parametrized by its $\bcx$ eigenvalue, each \emph{filled} state at $X$ must continuously interpolate, in the direction of decreasing $k_x$, to a state at $\bk$; this interpolation occurs in the same branch of $\bcx$, but the final state at $\bk$ \emph{may or may not be filled}. If $N_{\sma{+,\bk}} =N_{\sma{-,\bk}}=f/2$, there are as many odd/even states (at $\bk$) as there are odd/even states (at $X$) -- then it is possible for the interpolation to occur entirely among the filled states, as exemplified by Fig.\ \ref{fig:bandinversion}(a).  However, if $N_{\sma{+,\bk}} > N_{\sma{-,\bk}}$, then a number $(N_{\sma{+,\bk}}-N_{\sma{-,\bk}})/2$ of filled states (at $X$) must interpolate  to unfilled states (at $\bk$), giving rise to $(N_{+,\Gamma}-N_{-,\Gamma})/2$ chiral modes in the even branch of $\bcx$, as illustrated for $N_{\sma{+,\Gamma}}-N_{\sma{-,\Gamma}}=1$ and $2$, respectively, in Fig.\ \ref{fig:bandinversion}(c) and (e). By a similar demonstration, the same number of odd, antichiral modes must interpolate between filled states at $\bk$ to unfilled states at $X$. This leads to minimally $(N_{\sma{+,\bk}}-N_{\sma{-,\bk}})/2$ screw-protected, Dirac crossings between chiral and antichiral modes, as highlighted by blue squares in Fig.\ \ref{fig:bandinversion}(c) and (e). If instead $N_{\sma{+,\bk}}  <  N_{\sma{-,\bk}}$, an analogous demonstration leads to minimally $(N_{\sma{-,\bk}}-N_{\sma{+,\bk}})/2$ Dirac crossings [e.g., Fig.\ \ref{fig:bandinversion}(b)]. We have qualified $|N_{\sma{+,\bk}}-N_{\sma{-,\bk}}|/2$  as the \emph{minimal} Dirac count ($\bar{D}_{\bk}$), because the total count ($D_{\sma{\bk}}$) can in principle be greater than $\bar{D}_{\bk}$) by any positive even number,   
due to band inversions \emph{away} from $\bk$, e.g., Fig.\ \ref{fig:bandinversion}(d) illustrates two Dirac crossings along $\Gamma X$ (hence, $D_{\sma{\Gamma}}=2$) despite $N_{+,\Gamma}=N_{-,\Gamma}$.  This completes our proof of \q{genD} for $\bk$ along $X\Gamma X$; the proof for $\bk$ along $MYM$ is obtained by cosmetically substituting $\Gamma \rightarrow Y$ and $X \rightarrow M$ in the above demonstration. 

Since the above results depend essentially on the integer-spin representation of time-reversal and screw symmetries, they (and the topological characterization described in the next Section) would also apply to intrinsically spinless systems such as photonic crystals,\cite{topologicalphotonics,Weyldiscovery_Lu} though certain terms that naturally describe Fermi systems have to be re-interpreted. While `filling' is conventionally associated with Pauli exclusion, we may, at each wavevector, distinguish between `filled' and `unfilled' photonic bands separated by a frequency gap. There is, of course, no photonic `semimetal' in the sense of charge transport, though we may still discuss Dirac-type touchings between `filled' and `unfilled' bands.

\subsection{Topological characterization of band inverted semimetals} \la{sec:top}

Each screw-protected Dirac touching is associated with a $\pi$ quantized Berry phase, which is acquired in traversing a screw-symmetric momentum loop around the Dirac node. More generally, we consider the parallel transport of filled Bloch waves around a momentum loop $l$, where at each $\bk  \in  l$ a spectral gap separates a set of lower-energy, filled states (numbering $f$) from a higher-energy, unfilled subspace. The $f$-by-$f$ matrix representing such parallel transport is known as the Wilson loop,\cite{wilczek1984} and it may be expressed as the path-ordered exponential (denoted by $\overline{\text{exp}}$) of the Berry-Wilczek-Zee connection\cite{wilczek1984,berry1984} $\boldsymbol{\bA}(\boldsymbol{k})_{ij} = \langle {u_{i,\boldsymbol{k}}} |{\nabla_{\boldsymbol{k}}u_{j,\boldsymbol{k}}} \rangle$: 
\begin{align} \label{wloopdifferentiable}
\W[l] = \overline{\text{exp}}\,\left[{-\int_l} d\bk \cdot \boldsymbol{A}(\boldsymbol{k})\,\right],
\end{align}
where $\ket{u_{j,\boldsymbol{k}}}$ is an occupied eigenstate of the tight-binding Hamiltonian. The $U(f)$ Berry gauge field ($\bA$) may be decomposed into trace-ful and trace-less components, where the trace-ful term (Tr$[\bA]$) generates the Abelian component of the parallel transport:
\e{ e^{i\Phi_{\sma{U(1)}}[l]}\equiv\text{exp}\left[{-\int_l} d\bk \cdot \text{Tr}[\,\boldsymbol{A}(\boldsymbol{k})\,]\,\right] = \text{det}\big[\,\W[l]\,\big]. \la{u1berry}} 
The $U(1)$ Berry phase ($\Phi_{\sma{U(1)}}[l]$) is quantized to $\pi$ if we choose $l$ to encircle the Dirac node screw-symmetrically. By a `screw-symmetric circle', we mean that $l$ [exemplified by $l_1$ and $l_2$ in \fig{fig:wilsonloops}(c)] is contractible, and is mapped to $-l$ by $\bcx$, where the sign of $l$ indicates its orientation; the mapping follows from $\bcx: y\rightarrow-y$ in real space, and therefore $k_y\rightarrow-k_y$ in momentum space. More generally, given any symmetry ($g$) that maps $l\rightarrow -l$ and has a unitary representation ($\breve{g}^{\sma{-1}}=\dg{\breve{g}}$), then\cite{AA2014} 
\e{\breve{g}\,\W[l]\,\breve{g}^{-1} =\W[-l]=\W[l]^{-1} \;\Rightarrow\; \text{det}[\,\W[l]\,]=\pm 1.\la{quantize}}
The left equation implies that the spectrum of $\W[l]$ is invariant under complex conjugation; the right equation describes the  quantization of the $U(1)$ Berry phase, which is robust against any deformations of the Hamiltonian that preserve both the  symmetry ($g$) and the spectral gap along $l$. In our context, $g=\bcx$, and exp$(i\Phi_{\sma{U(1)}}[l])=1$ (resp.\ $-1$) if $l$ encircles an even (resp.\ odd) number of Dirac fermions.

Our topological discussion has thus far focused on characterizing \emph{individual} Dirac nodes, by an \emph{Abelian} Berry-type invariant defined over \emph{contractible} momentum loops; a global characterization of \emph{all} Dirac nodes is possible with a \emph{non-Abelian} Berry invariant\cite{AA2014,Maryam2014} defined over a \emph{noncontractible}\cite{zak1989} momentum loop. By a non-Abelian Berry invariant, we mean that it requires knowledge of the individual eigenvalues of $\W[l]$, which encode the non-Abelian transport generated by the trace-less component of $\bA$.

Henceforth, we consider only screw-symmetric loops $l(k_x)$ parallel to $\bs{e}_y$ and at fixed $k_x$, as illustrated in \fig{fig:wilsonloops}(c); we thus shorten $\W[l(k_x)]$ to $\W(k_x)$. Applying \q{quantize} for $l(k_x)$, the invariance of the $\W(k_x)$-spectrum under complex conjugation implies that a (possibly zero) subset of $\W(k_x)$-eigenvalues (numbering $W_{\pm}(k_x)$) is respectively quantized to $\pm 1$. In App.\ \ref{app:wilson}, we relate these quantized $\W(k_x)$-eigenvalues to the total number (${D}_{\sma{l(k_x)}}$) of Dirac crossings in the cylinder bounded by $l(k_x)$ and $MXM$ [red-shaded region in \fig{fig:wilsonloops}(c)]:
\e{ {D}_{\sma{l(k_x)}} \eq \text{max}\{W_+(k_x),W_-(k_x)\}+2b; \; b\in \Z^{\sma{\geq}}. \la{wilsondirac3}}
As described in \s{sec:bandinversion}, such crossings can originate from independent band inversions on either screw intervals to the right of $l(k_x)$ [marked as dashed lines in \fig{fig:wilsonloops}(c)]. The difference in Dirac counts between the two intervals also relates to the quantized $\W$-eigenvalues as  
\begin{equation}
{D}_{\sma{(k_x,0)}}-{D}_{\sma{(k_x,\pi)}} - \text{min}\{W_+(k_x),W_-(k_x)\} \in 2\Z, 
\label{wilsondirac4}
\end{equation}
where $2\Z$ denotes the set of even integers. Additional contributions to ${D}_{\sma{l(k_x)}}$ may arise from screw-symmetric pairs of crossings away from screw lines, as may be stabilized by a spatial symmetry (e.g., inversion) other than screw.

We exemplify our result for crystals with fillings $2$ and $4$; their possible screw representations, $\W$-spectra and minimal Dirac counts are tabulated in Table\ \ref{table1d2band} (in this Section) and \ref{table1d4band} (in App.\ \ref{app:wilson}). These properties are then applied to topologically distinguish the different phases in Fig.\ \ref{fig:bandinversion}(a-e), which are all assumed to have no Dirac crossings along the unillustrated screw line $YM$, and also no crossings away from the screw lines. We exemplify this analysis for  \fig{fig:bandinversion}(b), focusing on the interval $0 \leq k_x < k_x'$. The filled bands are inverted in this segment  of $\Gamma X$, with $N_{\sma{+,(k_x,0)}}=2, N_{\sma{-,(k_x,0)}}=0$; by assumption of $YM$, $N_{\sma{+,(k_x,\pi)}}=N_{\sma{-,(k_x,\pi)}}=1$. Thus reading off the third-from-bottom row in Table\ \ref{table1d2band}, we obtain $W_{\pm}(k_x)=1$ (illustrated in the bottom-most plot of \fig{fig:bandinversion}(b)). This further implies from \q{wilsondirac3} that the total Dirac count ${D}_{\sma{(k_x,0)}}+{D}_{\sma{(k_x,\pi)}}=1+2c$ with $c \in \Z^{\sma{\geq}}$, and from \q{wilsondirac4} that both ${D}_{\sma{(k_x,0)}}$and ${D}_{\sma{(k_x,\pi)}}$ have the same parity; \fig{fig:bandinversion}(b) shows in fact that ${D}_{\sma{(k_x,0)}}=1$ and by assumption ${D}_{\sma{(k_x,\pi)}}=0$. Finally, we remark that the Wilson loop in \q{wloopdifferentiable} is only well-defined and continuous on intervals that exclude the Dirac points, e.g., the discontinuity of $W_{\pm}(k_x)$ at $k_x'$ in \fig{fig:bandinversion}(b) necessarily indicates a Dirac crossing. Our schematic example in \fig{fig:bandinversion}(b) is further materialized by the spin-orbit-free \wt~monolayer, as we demonstrate in Sec.\ \ref{sec:tbspinless}.

\begin{table}[t]
	\centering
		\begin{tabular} {|c|c|c|c|c|} \hline
			$N_{\sma{+,\bk_1}}$  & $N_{\sma{-,\bk_2}}$ &$\W(k_x)$ & $\bar{D}_{\sma{l(k_x)}}$ &  ${D}_{\sma{\bk_1}}-{D}_{\sma{\bk_2}}$ mod $2$ \\  \hline \hline 
		  1&1 & $[\lambda_1 \lambda_1^{\ast}]$& 0&0\\ \hline
			 2&1 & $[+ -]$&  1 & 1\\ \hline
			 2&0 & $[- -]$& 2 & 0\\ \hline
			 2&2 & $[+ +]$& 2 & 0\\ \hline
		\end{tabular}
		\caption{Topological characterization of a nonsymmorphic crystal with filling $f=2$ and either screw or glide symmetry. $\bk_1$ and $\bk_2$ are shorthand for the two screw-invariant (or glide-invariant) momenta at a particular $k_x$: $(k_x,0)$  and $(k_x,\pi)$. Leftmost column: $N_{\sma{+,\bk_1}}$ is number of filled states at $\bk_1$ in the even screw (glide) branch, i.e., with screw (glide) eigenvalue $+$exp$(-ik_x/2)$; $N_{\sma{-,\bk_1}}=f-N_{\sma{+,\bk_1}}$ is the number of odd, filled states with screw (glide) eigenvalue $-$exp$(-ik_x/2)$. Second column from the left: number of even filled states at $\bk_2$. Third column: $\W(k_x)$-spectrum, with $+$($-$) representing a $+1$ ($-1$) eigenvalue, and $\lambda_i\lambda_i^*$ a unimodular, complex-conjugate pair. Fourth column: the minimal number (${\equiv}\bar{D}_{\sma{l(k_x)}}$) of Dirac crossings in a momentum cylinder bounded by $l(k_x)$ and $MXM$.  Rightmost column: the parity of the difference in Dirac counts along either screw intervals. We remark that the $\W(k_x)$-spectrum depends only on relative changes in the symmetry representations between $\bk_1$ and $\bk_2$, i.e., it is invariant under: (i) interchanging $N_{\sma{+,\bk_1}}$ with $N_{\sma{+,\bk_2}}$, and (ii) multiplication of all screw (glide) eigenvalues by a common factor $-1$, i.e., $\{N_{\sma{+,\bk_1}},N_{\sma{+,\bk_2}}\} \rightarrow \{N'_{\sma{+,\bk_1}}=f-N_{\sma{+,\bk_1}},N'_{\sma{+,\bk_2}}=f-N_{\sma{+,\bk_2}}\}$. For example, the $\W(k_x)$-spectrum and Dirac counts in the third-to-last row additionally describe a crystal with two odd states at $\bk_1$ and an odd/even pair at $\bk_2$, and also describe a crystal with two odd states at $\bk_2$ and an odd/even pair at $\bk_1$.}
		\label{table1d2band}
\end{table}

Given the well-known $U(1)$ ambiguity of the Wilson loop originating from the choice of real-spatial origin,\cite{AA2014}  all \emph{other} higher-than-one-dimensional band topologies (with one exception\cite{Cohomological}) are diagnosed by gauge-invariant {differences} of Wilson loops, between different 1D submanifolds in the same Brillouin zone (BZ).\cite{fu2006,AA2014,berryphaseTCI,alexey2011,yu2011,Maryam2014}  On the other hand, we propose that the invariants in \q{wilsondirac3} and (\ref{wilsondirac4}), as extracted from a single Wilson loop, are predictive of the Dirac semimetallicity in a 2D submanifold of the Brillouin zone -- since the existence of Dirac points cannot depend on the choice of origin, our invariants are likewise independent, as we proceed to demonstrate. Since the Wilson loops we consider traverse a momentum path which is parallel to $\bs{e}_y$, their eigenspectrum only depends on the $y$-coordinate of the origin. Specifically, translating the spatial origin by $\delta \br$ induces a global phase shift of all $\W$-eigenvalues by $\bG {\cdot} \delta \br$, with $\bG$ the reciprocal period along $\bs{e}_y$.\cite{AA2014} In a 2D Bravais lattice, there are always two inequivalent, real-spatial lines which are invariant under the screw rotation, as exemplified by horizontal dashed lines in \fig{fig:wilsonloops}(a) for the \wt~monolayer; assuming no spatial symmetry other than screw, the spatial origin may lie at any point on either screw line. Since the two lines are separated by half a lattice period in $\bs{e}_y$, translating the origin between these lines induces a global phase shift of $\bG \cdot \delta \br=\pm\pi$. Consequently, $W_{\pm} \rightarrow W_{\mp}$, but their maximum and minimum values in \q{wilsondirac3} and (\ref{wilsondirac4}) are clearly invariant. We call any such quantity, that is both extractable from a single Wilson loop and insensitive to the spatial origin, a strong Wilson-loop invariant; another known example classifies a newly-introduced nonsymmorphic topological insulator;\cite{Cohomological} all other known single-Wilson-loop invariants\cite{zak1989,AA2014,berryphaseTCI} are comparatively weak, and relate to spatially-dependent physical predictions. A case in point is the geometric theory of polarization,\cite{kingsmith1993} which predicts different electronic charges at the edge of a crystal,  depending on where the edge is terminated.\cite{vanderbilt1993}

\subsection{Proximity of topological semimetals to the $\Z_2$-topological insulator}\la{sec:qshi}

Topological semimetals are often linked to gapped, topological phases. There are various ways to arrive at such a gapped phase: (i) the mutual annihilation of band crossings with zero net topological charge, (ii) the breaking of a spatial symmetry, and (iii) the introduction of spin-orbit coupling in electronic semimetals,\cite{Young} which may also be interpreted as a breaking of spin-$SU(2)$ symmetry. To exemplify (i-iii) in this order, inversion-asymmetric Weyl semimetals intermediate between a trivial, gapped phase and a $\Z_2$-topological, gapped phase;\cite{Murakami2007B} certain 3D Dirac semimetals\cite{Na3BiZhijun} are gapped when their protective spatial symmetry is broken, leading to a novel, nonsymmorphic topological phase;\cite{GSHI} the slightest spin-orbit coupling gaps graphene,\cite{kane2005A} a symmorphic Dirac semimetal, to form a quantum spin Hall phase with $\Z_2$ topological order\cite{kane2005B,QSHE_Rahul,QSHE_bernevig} (in short, a $\Z_2$-topological insulator). 

\begin{figure}[t]
\centering
\includegraphics[width=8 cm]{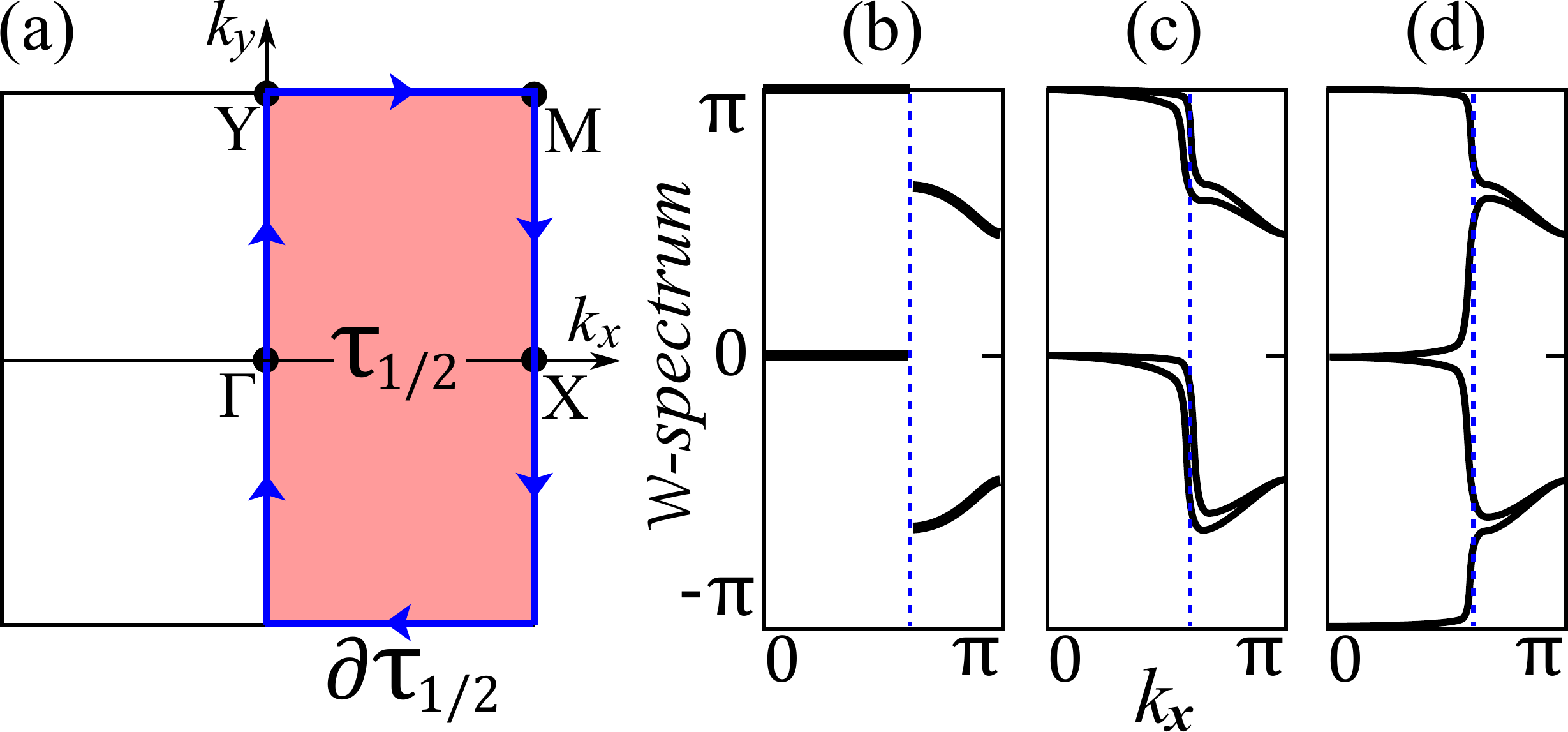}
\caption{ (b-d) is a pictorial argument for a quantum spin Hall phase, given a semimetal with a single Dirac fermion (per spin component) in $\tau_{\sma{1/2}}$ (red-shaded half Brillouin zone in (a)).}\label{fig:qshi}
\end{figure}

This Section describes how, analogously to graphene, the slightest spin-orbit coupling gaps a nonsymmorphic Dirac semimetal to form a $\Z_2$-topological insulator. Not all Dirac semimetals necessarily lead to a $\Z_2$-topological phase when gapped -- the criterion on the semimetal may be stated in three equivalent ways. If and only if the nonsymmorphic semimetal is characterized by:\\
\noi{i} an odd number of Dirac fermions (per spin component)  in half the Brillouin zone  [red-shaded region labelled by $\tau_{\sma{1/2}}$ in \fig{fig:qshi}(a)], or, equivalently, \\
\noi{ii} the Abelian component of the single-spin Wilson loop along $Y\Gamma Y$ (i.e., det$[\W(0)]$) equals $-1$, or, equivalently,\\
\noi{iii} the product of the nonsymmorphic eigenvalues (whether glide or screw) of single-spin filled states over $\Gamma$ and $Y$ equals $-1$,\\ 
\noindent then any gap-inducing, spin-orbit coupling that preserves the nonsymmorphic symmetry (whether glide or screw) results in a $\Z_2$-topological phase. 

More precisely, the above criterion guarantees the $\Z_2$-topological phase for weak spin-orbit coupling; in principle, one cannot rule out that strong spin-orbit coupling might induce a transition to a trivial, gapped phase. That (i) and (ii) are equivalent follows from \q{wilsondirac3}, where we deduce that odd $D_{\sma{l(0)}}$ is equivalent to either odd $W_+(0)$ or odd $W_-(0)$. Due to the assumed-even filling and the invariance of the $\W(0)$-spectrum under complex conjugation (cf.\ \q{quantize}), it is always the case that if either is odd, both $W_+(0)$ and $W_-(0)$ are odd, which leads to det$[\W(0)]=-1$. That (i) and (iii) are equivalent follows from \q{genD} which leads to
\e{  
D_{l(0)}\sim D_{\Gamma}+D_{Y} \sim \f{|N_{+,\Gamma}-N_{-,\Gamma}|}{2}+\f{|N_{+,Y}-N_{-,Y}|}{2}, \notag
}
where $\sim$ denotes an equality modulo two. On further application of $f=N_{\sma{+,\bk}}-N_{\sma{-,\bk}} \in 2\Z$, we deduce that odd $D_{l(0)}$ occurs iff $N_{\sma{-,\Gamma}}+N_{\sma{-,Y}}$ is also odd; finally, observe that the product of nonsymmorphic eigenvalues equals $(-1)^{\sma{N_{\sma{-,\Gamma}}+N_{\sma{-,Y}}}}$. We remark that (iii) is the nonsymmorphic generalization of the Fu-Kane criterion\cite{Inversion_Fu} for a $\Z_2$-topological phase in centrosymmetric crystals.  

To pictorially argue for our criterion, we return to our semimetallic case study in \fig{fig:bandinversion}(b), which has only a single Dirac fermion along $\Gamma X$ and no other Dirac crossing in the cylinder bounded by $Y\Gamma Y$ and $MXM$. The filled states are characterized by $N_{\sma{-,\Gamma}}=2, N_{\sma{+,\Gamma}}=0$ and $N_{\sma{+,Y}}=N_{\sma{-,Y}}=1$, which implies that the product of their screw eigenvalues equals $-1$. The third-from-bottom row of Table\ \ref{table1d2band} informs us that the two $\W(0)$-eigenvalues (per spin component) are $+1$ and $-1$, which implies that det$[\W(0)]=-1$. For $0 \leq  k_x  \leq  \pi$, the spectrum of $\W(k_x)$  is reproduced in \fig{fig:qshi}(b), where each Wilson `band' is spin-degenerate and discontinuous at the momentum position of the Dirac fermion (indicated by the blue dashed line). The introduction of spin-orbit coupling splits this spin degeneracy everywhere except for the Kramers degeneracies at the time-reversal-invariant $k_x  \in \{0,\pi\}$. Additionally, the gapping of the Dirac fermion implies  that a subspace of filled states is now smoothly defined over the entire Brillouin zone -- this smoothens out the discontinuity in the $\W(k_x)$-spectrum. If we only assume that the spin-orbit coupling is time-reversal symmetric, there are two ways of smoothening: \fig{fig:qshi}(c) (resp.\ (d)) illustrates the Kramers-partner-preserving (resp.\ partner-switching\cite{fu2006,yu2011,alexey2011}) doublets of the trivial gapped phase (resp.\ the $\Z_2$-topological phase). If we further assume that the spin-orbit coupling respects the nonsymmorphic symmetry (whether glide or screw), then the $\W(k_x)$-spectrum is further constrained to be invariant under complex conjugation (cf.\ \q{quantize}), which uniquely selects the $\Z_2$-topological phase of \fig{fig:qshi}(d). Beyond this pictorial argument, a technical proof of our criterion is provided in App.\ \ref{app:qshi}.

Our criterion is more broadly predictive of $\Z_2$-topological phases which are semimetallic from the perspective of transport, so long as a finite energy difference exists between two sets of bands for all $\bk$ -- the time-reversal $\Z_2$ invariant\cite{kane2005B} is then well-defined for both sets of bands. This qualifier is relevant to spin-orbit-coupled WTe$_2$ monolayers, which have $\Z_2$ topological order in conjunction with electron and hole pockets, as we demonstrate in \s{sec:spinfulmono}. 

\section{WT\lowercase{e}$_2$ Monolayer}
\label{sec: mono}

In \s{sec:tbspinless}, we present a minimal tight-binding model of a spin-orbit-free \wt~monolayer, which confirms its topological semimetallicity in the sense of \s{sec:bandinversion}. This model also captures the tilting of the Dirac fermion, which we formalize in \s{sec:typeII} by introducing the notion of a type-II Dirac fermion. Finally in \s{sec:typeII}, we apply our criterion in \s{sec:qshi} to predict that spin-orbit-coupled \wt~monolayer has $\Z_2$-topological order.

\subsection{Tight-binding model and topological characterization of the spin-orbit-free $\mathrm{WTe}_2$ monolayer}\la{sec:tbspinless}

\begin{figure}[t]
 \centering
 \includegraphics[width=0.95\columnwidth]{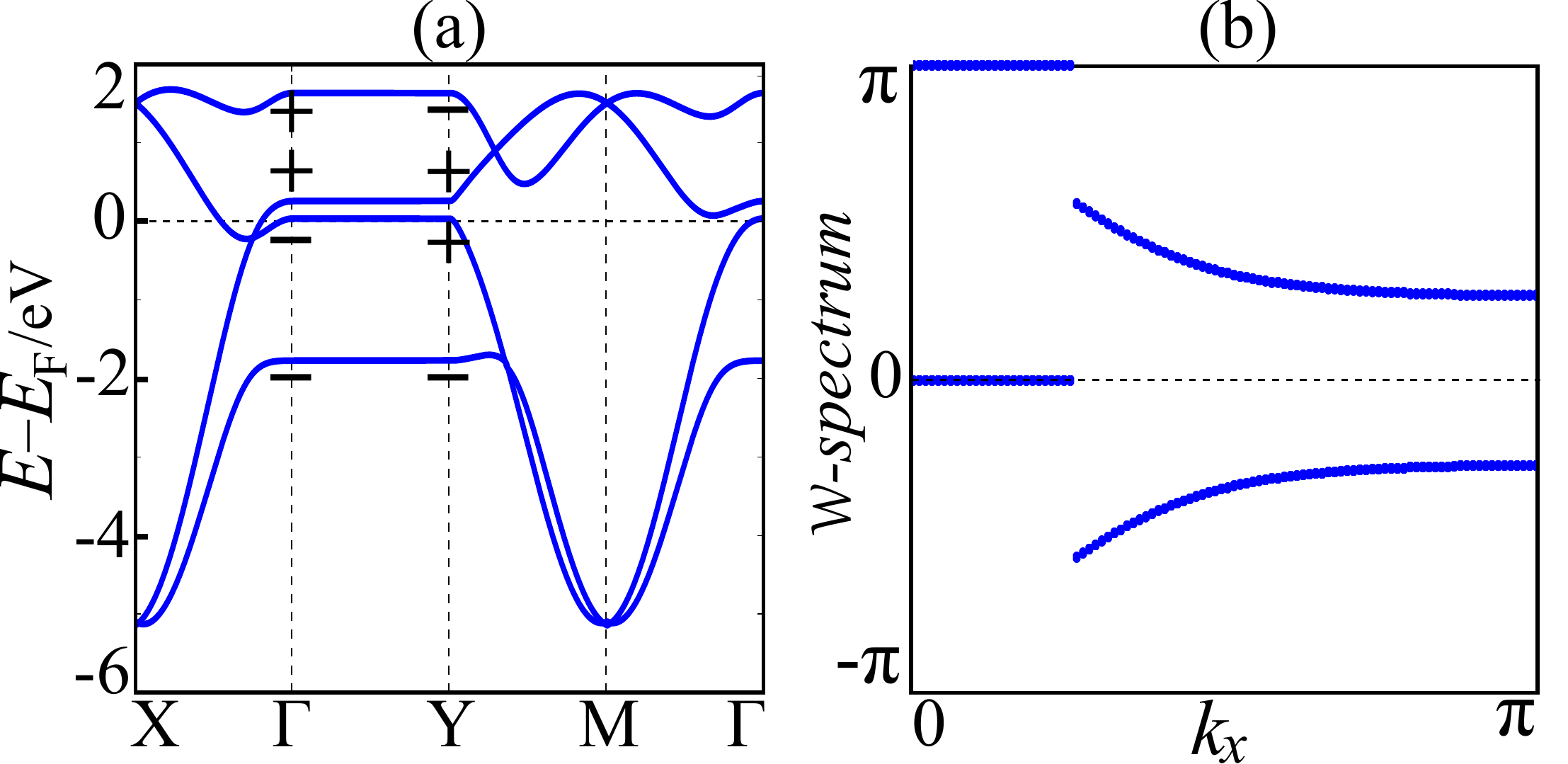}
 \caption{(a) Bandstructure for the tight-binding model [\q{tbmod}] of spin-orbit-free \wt, with parameters listed in App.\ \ref{app: tight-binding}. At $\Gamma$ and $Y$, $+$ and $-$ respectively label bands with screw-eigenvalues $+1$ and $-1$. (b) Phases of the Wilson-loop eigenvalues, as calculated for the same tight-binding model [\q{tbmod}].}
\label{fig_6}
\end{figure}

We now present a minimal tight-binding model of the spin-orbit-free \wt~monolayer, which reproduces the DFT bandstructure near the Fermi level; compare \fig{fig_6}(a) with \fig{fig_2}(c). Our model includes the minimal number (four) of bands to describe a band inversion between two sets of two-fold-connected bands: cf. \s{sec:comparingsemimetals}. With modified tight-binding parameters, this model can be more generally be applied to MX$_2$ compounds with the same crystal structure. 

By a Wannier interpolation\cite{wannier} of these four DFT bands, we construct a basis of maximally-localized Wannier functions, comprising two ${d}_{x^2-y^2}$-type orbitals which derive from the W atoms [indicated by $M$ in Fig.~\ref{fig:wilsonloops}(a)], and two $p_x$-type orbitals derived from a subset of the Te atoms [$X$-2 in Fig.~\ref{fig:wilsonloops}(a)]; there are no orbitals derived from the complementary Te sublattice [$X$-1 in Fig.~\ref{fig:wilsonloops}(a)] in our low-energy description. As indicated within the rectangular unit cell of \fig{fig:wilsonloops}(a), the centers of these Wannier functions divide into two sublattices, labelled by $A$ and $B$, which are permuted by the screw transformation $\bcx$. Each ${d}_{x^2-y^2}$-type (\ ${p}_{x}$-type) Wannier function is even (odd) under a mirror operation $x \mapsto -x $ centered at the W atom (Te atom).
 
In this reduced Hilbert space, our tight-binding Hamiltonian includes all symmetry-allowed, nearest-neighbor hoppings, as well as two next-nearest-neighbor, intra-sublattice hoppings along the chain: 
\begin{equation} 
\begin{split}\la{tbmod}
H = \sum_{\bs{R}} \bigg[
&\phantom+ \sum_{\ell}  \mu_\ell \left( a^\dagger_{\ell,\bs{R}} a^{\phantom{\dagger}}_{\ell,\bs{R}}  + b^\dagger_{\ell,\bs{R}} b_{\ell,\bs{R}}^{\phantom{\dagger}}   \right) \\
&+\sum_{\ell} t_\ell \left(
 a^\dagger_{\ell,\bs{R} + \bs{e}_x } a_{\ell,\bs{R}}^{\phantom{\dagger}}  +  
b^\dagger_{\ell,\bs{R} + \bs{e}_x } b_{\ell,\bs{R}}^{\phantom{\dagger}} \right) \\
&+\sum_{n=0}^1 (-1)^nt_0^{AB} \big(  b^\dagger_{p,\bs{R} + n \bs{e}_x} a^{\phantom{\dagger}}_{d,\bs{R}}- b^\dagger_{d,\bs{R} + n \bs{e}_x} a^{\phantom{\dagger}}_{p,\bs{R}} \big) \\
&+\sum_{\ell} \sum_{\bs{\delta}_\ell} t_l^{AB} a^\dagger_{\ell,\bs{R}+\bs{\delta}_\ell} b_{\ell,\bs{R}}^{\phantom{\dagger}} 
\bigg] + \text{h.c.}.
\end{split}
\end{equation}
Here, $\ell$ denotes the orbital type  (either $p$ or $d$);  $a^\dagger_{\ell,\bs{R}}$ ($b^\dagger_{\ell,\bs{R}}$) creates an $\ell$-type Wannier function centered at the position $r_{\ell,A}+\bs{R}$ (resp.\ $r_{\ell,B}+\bs{R})$ as indicated in Fig.\ \ref{fig:wilsonloops}(a); $\bs{\delta}_{\ell}$ are vectors given by $\bs{\delta}_p = \bs{0},\bs{e}_x$ and $\bs{\delta}_d = \bs{e}_x+\bs{e}_y,\bs{e}_y$. Since our Hamiltonian is spin-$SU(2)$ symmetric, we omit the spin label for each electron operator. The tight-binding parameters are listed in App.\ \ref{app: tight-binding}, where we also express \q{tbmod} in the momentum representation.

For
\begin{equation}
t_p - t_d  < \frac{\mu_p - \mu_d}{2} + \left( t_p^{AB}+ t_d^{AB}  \right),
\end{equation}
orthogonal screw representations at $\Gamma$ are inverted (i.e., $0 =N_{\sma{+,\Gamma}} \neq N_{\sma{-,\Gamma}} = 2$),
as we illustrate in Fig.\ \ref{fig_6}(a) for WTe$_2$. This band inversion  leads to a time-reversed pair of Dirac fermions along the screw line $X\Gamma X$, which are encoded as a discontinuity in the Wilson-loop spectra in Fig.\ \ref{fig_6}(b). 
These Dirac nodes necessarily belong to the screw line but they can have any position along it, i.e. by perturbing the Hamiltonian we might move the Dirac nodes along $X\Gamma X$.



\subsection{Type-II Dirac fermions}\la{sec:typeII}

The Dirac crossing along the screw line ($\Gamma X$) is described by an effective Hamiltonian: 
\begin{equation}
\mathcal{H}_{\mathrm{II}}(\bs{k}) = u_x k_x \sigma_0 + v_x k_x \sigma_1 + v_y k_y \sigma_2,
\label{eq: type-II Dirac equation} 
\end{equation}
to linear order in the momentum coordinates originating from the Dirac node, which we define as the point of degeneracy; $\sigma_{1,2}$ are Pauli matrices, and $\sigma_0$ the identity matrix, in a pseudospin basis labelled by $\sigma_1\ket{\pm}=\pm \ket{\pm}$, with the orbital characters 
$\ket{\pm} = i \alpha_{\pm} \ket{p_{\pm}} + \beta_{\pm} \ket{d_{\pm}},$ 
where $\ket{l_{\pm}} =  \left( \ket{l,A} \pm \ket{l,B} \right)/\sqrt{2}$ are the bonding/anti-bonding combinations of $l$-type orbitals across the two sublattices. The Greek symbols are real, normalized according to $\alpha_{\pm}^2 + \beta_{\pm}^2 = 1$, and determined numerically by diagonalizing \q{tbmod} at the Dirac node.

The group ($G$) of this nodal wavevector\cite{tinkhambook} is generated by the screw $\bcx$ and the space-time inversion $T\cali$ (i.e., the product of time reversal $T$ and spatial inversion $\cali$), which respectively transform the Hamiltonian as 
\e{&\sigma_1\calh_{\mathrm{II}}(k_x,k_y)\sigma_1 = \calh_{\mathrm{II}}(k_x,-k_y),\lin
&\sigma_1\calh_{\mathrm{II}}^*(k_x,k_y)\sigma_1 = \calh_{\mathrm{II}}(k_x,k_y). \la{spaco}}
The lack of any discrete rotational symmetry in $G$ implies that the spectrum of $\calh_{\mathrm{II}}$ is rotationally anisotropic, as evidenced by
\e{\epsilon_{\pm}(\bs{k}) = u_x k_x \pm \sqrt{v_x^2 k_x^2  + v_y^2 k_y^2}.}
The resulting anisotropy of the Fermi velocities may have important consequences in transport. This is in contrast to graphene, where each Dirac cone is fixed to a wavevector that is invariant under three-fold rotation, such that the Dirac spectrum of graphene has an emergent, continuous-rotational symmetry. 

For \wt, $G$ also lacks the reflection symmetry that maps $k_x \rightarrow -k_x$, as manifested by the allowed term $u_x k_x \sigma_0$ in the Hamiltonian. This term induces a tilting of the Dirac cone, which originates from the intra-sublattice hoppings $t_\ell$ in \q{tbmod}. If the tilting is sufficiently pronounced, part of the upper Dirac cone dips below the nodal energy [$\epsilon_{\pm}(\bs{0})=0$] as illustrated in \fig{fig_1}(a), resulting in a discontinuous change in the band contours at the nodal energy, i.e., a Lifshitz transition between a type-I ($|u_x| < |v_x|$) spectrum with a closed Fermi circle surrounding $\bk=0$, and a type-II ($|u_x| >|v_x|$) spectrum with open Fermi line(s) to linear order in $\bk$; the latter case applies to WTe$_2$, as we illustrate in \fig{fig_1}(a-b).\cite{WTe2Weyl,volovik} All lattice-regularized Fermi surfaces are of course closed when higher-order momenta are accounted for -- the regularized, type-II spectrum is described by two Fermi circles, with one being hole-type and the other electron-type. If the  Fermi level lies at the type-II Dirac node (i.e., $\varepsilon_F = \epsilon_{\pm}(\boldsymbol{0})=0$), the electron- and hole-type Fermi circles touch at the Dirac node. However, there is no symmetry constraint on the Fermi energy and so both circles are generically disconnected; for negative $\varepsilon_F$ (resp.\ positive $\varepsilon_F$, as applies to WTe$_2$), the hole-type (resp.\ electron-type) Fermi circle is characterized by a $U(1)$-Berry phase [$\Phi_{\sma{U(1)}}$ in \q{u1berry}] equal to $\pi$, while the electron-type (resp.\ hole-type) Fermi circle has $U(1)$-Berry phase equal to $0$; this quantization is proven in \s{sec:top}, where we would show that each Fermi circle corresponds to a screw-symmetric quasimomentum loop.



\subsection{Spin-orbit-coupled WT\lowercase{e}$_2$ monolayer}\la{sec:spinfulmono}

The screw eigenvalues of the bands at $\Gamma$ and $Y$ are indicated in \fig{fig_6}(a); their product over the filled bands equals $-1$, which implies, through criterion (iii) in \s{sec:qshi}, that weakly-spin-orbit-coupled \wt~has $\Z_2$ topological order. This result has consistently been derived\cite{qian2014quantum} by exploiting the Fu-Kane criterion\cite{Inversion_Fu} for centrosymmetric crystals; we remark that our proposed criterion in \s{sec:qshi} more generally applies to nonsymmorphic semimetals without spatial-inversion symmetry.

\section{Bilayer}
\label{sec: bi}
\begin{figure}[t]
 \centering
 \includegraphics[width=0.95\columnwidth]{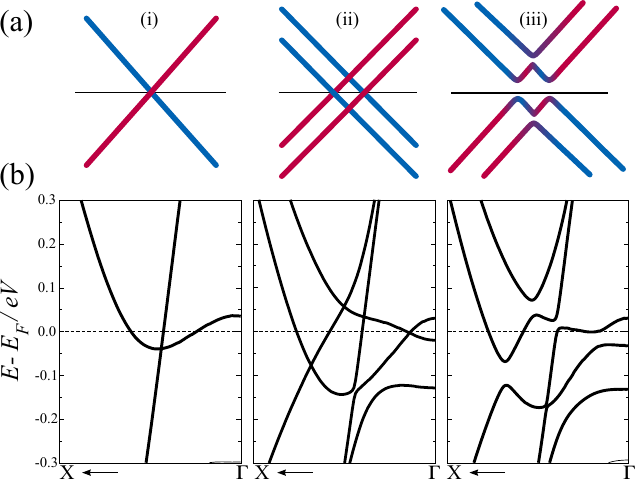}
 \caption{ (a) Hypothetical coupling of two band-inverted monolayers with type-I Dirac fermions: (a-i) depicts a doubly-degenerate Dirac crossing with vanishing coupling, (a-ii) a semimetallic state with two distinct Dirac cones if the nonvanishing coupling preserves the screw symmetry, (a-iii) a trivial insulator in the case of nonvanishing screw-symmetric coupling. (b) DFT band structures of bilayer \wt~for different inter-monolayer couplings: (b-i) vanishing, (b-ii) screw-symmetry-preserving, and (b-iii) screw-symmetry-breaking. }
\label{fig_7}
\end{figure}

We propose to interpret 3D WTe$_2$ as a periodic stacking of bilayers, for which the intra-bilayer couplings dominate over the inter-bilayer couplings.  We support our interpretation by finding that certain features of the 3D electronic structure, precisely the electron and hole pockets along the $\Gamma X$ line, are already present in the bilayer; we show in \s{sec:origin} how these pockets (in both bilayer and 3D) ultimately originate from the tilted Dirac cones of the single monolayer. In \s{sec:dichroism}, we further calculate the dichroism that originates from these pockets, so as to qualitatively explain the recently-measured CD in 3D WTe$_2$.\cite{Wte2CD} 

\subsection{Origin of electron and hole pockets in bilayer and 3D WT\lowercase{e}$_2$} \la{sec:origin}

The electron and hole pockets of 3D WTe$_2$ are a fundamental aspect of its electronic structure and are thought to be responsible for the large, non-saturating magnetoresistance. \cite{Wte2nature} Here, we propose that these pockets originate from the type-II Dirac fermions of the spin-orbit-free monolayer. 

Our argument is illustrated in \fig{fig_7}: to begin, in the limit of vanishing coupling between the monolayers, the bilayer bandstructure is doubly degenerate at all wavevectors. This follows because one monolayer is related to the other by a global continuous translation in real space, so that the energy-momentum dispersion curves of each monolayer are identical. We focus on a low-energy theory near the Dirac node, which is doubly-degenerate in the aforementioned limit. 

We now consider two different stackings for a bilayer of \wt: in the first, the two monolayers have no relative displacement in the $\bs{e}_x$-$\bs{e}_y$ plane but are displaced along $\bs{e}_z$ and therefore the bilayer retains the screw symmetry of the monolayer; we warn that this hypothetical stacking is unlikely to occur in nature. The inter-monolayer coupling energetically splits the two Dirac cones, which correspond to bonding and anti-bonding combinations of the two monolayer wavefunctions, as illustrated in \fig{fig_7}(b-ii); the degeneracy of the Dirac crossing is retained since screw symmetry is preserved. The low-energy theory of each Dirac crossing is described by the Hamiltonian in \q{eq: type-II Dirac equation}, albeit with slightly different parameters and basis wavefunctions. Independent of the particular basis, we note that \q{eq: type-II Dirac equation} describes a pseudospin that is coupled to a pseudomagnetic field ($\bB$) with components $B_x(\bk)=v_xk_x, B_y(\bk)=v_yk_y, B_z(\bk)=0.$ The vanishing of $B_z$ is a consequence of the spatial symmetries which constrain the Hamiltonian as in \q{spaco} -- the Hamiltonian eigenfunctions therefore correspond to a pseudospin that is confined to a pseudo $x-y$ plane, and moreover winds as we encircle the Dirac node. 

The second stacking corresponds to the experimental lattice parameters for 3D \wt, and breaks the screw symmetry, as we illustrate in \fig{fig:wilsonloops}(b). The asymmetric bandstructure differs from the symmetric stacking in that small gaps open at the Dirac nodes; compare \fig{fig_7}(b-ii) and (b-iii). Otherwise, both bandstructures are very similar away from the Dirac nodes, and in particular both possess electron and hole Fermi pockets in the vicinity of the Dirac nodes. Each screw-asymmetric Dirac fermion is described by a small mass term proportional to $\sigma_3$, i.e., $B_z$ is a nonvanishingly small constant.  Therefore, around either Fermi circle, the pseudospin rotates in the $x-y$ plane with a small out-of-plane component. In particular, the sense of rotation for the electron-like Fermi circle is opposite to that of the hole-like circle, with consequences for the dichroism that we elaborate in \s{sec:dichroism}. 

We argue that these Fermi pockets are not generic and originate from the tilting of the Dirac cone. For the sake of this argument, we schematically illustrate in \fig{fig_7}(a) a hypothetical bilayer which comprises two band-inverted monolayers. In this hypothetical scenario, the Dirac cone of the monolayer is type-I and the Fermi surfaces are point-like; therefore, screw-symmetry-breaking inter-monolayer couplings generically produce a fully-gapped, trivial insulator, as we argue pictorially through \fig{fig_7}(a).


\subsection{Dichroism} \la{sec:dichroism}
\begin{figure*}[t]
 \centering
 \includegraphics[width=2\columnwidth]{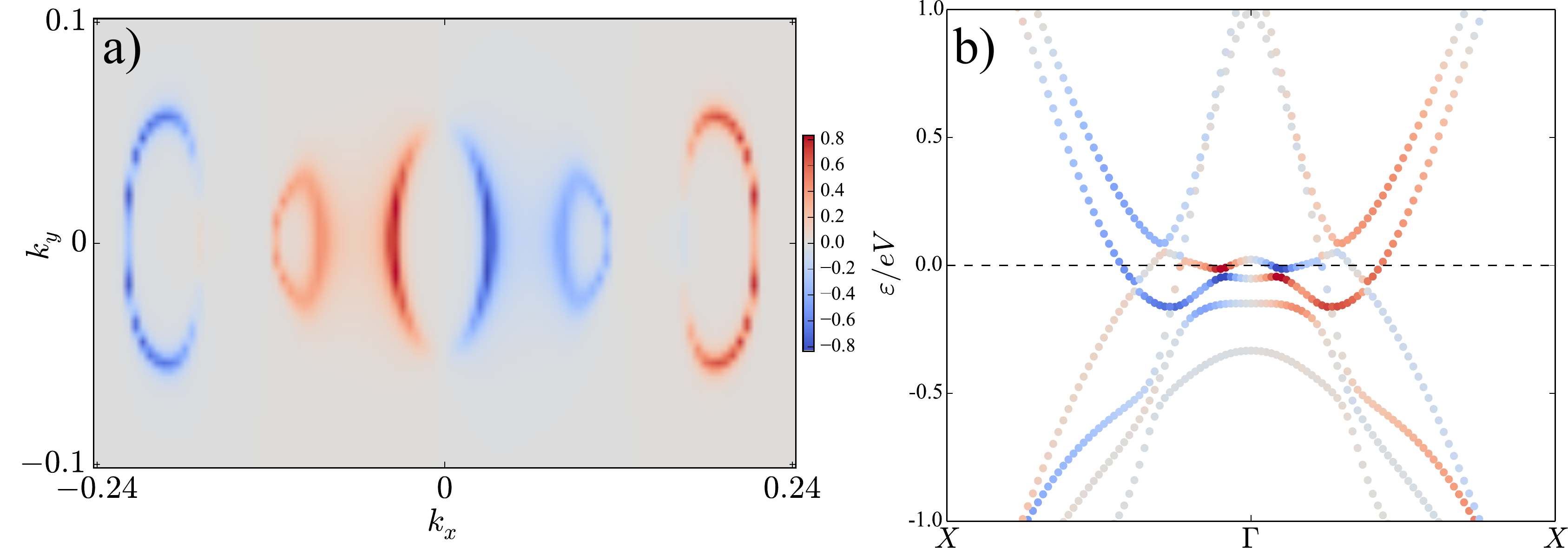}
 \caption{(a) Fermi-level dichroic signal [$D_s(\bs{k},\varepsilon_{\textrm{F}})$ defined in the main text] for a WTe$_2$ bilayer. The Fermi surface has been broadened with a Lorentzian with a full width at half maximum of 10 meV. The Fermi pocket around $\Gamma$\cite{WTe2arpes,Wte2CD,FS_calc_discussion_2} does not encircle a massive Dirac fermion.
 (b) $D_s(\bs{k},E)$ for the band structure along $X \Gamma X$ for energies near the Fermi level.}
\label{fig_8}
\end{figure*}
Circular dichroism (CD) has recently been observed in the angle-resolved photoemission spectrum (ARPES) of 3D WTe$_2$,\cite{Wte2CD} i.e., the photoemission is dependent on the helical polarization of light. CD depends on the wavefunction of the initial and of the final electronic state upon photon absorption and is therefore a sensitive probe of the electronic structure of a material. \cite{CDschoenhense} A simple and widely-applied model of photoemission breaks down the process into three steps: (i) photoexcitation of an electron in the solid, (ii) propagation of the photoelectron to the surface, and, (iii), escape of the photoelectron into the vacuum.~\cite{3step}
CD is encoded in the first step, where the photon causes a transition between two Bloch states in the solid. 
The intensity of the photoemitted electrons with wavevector $\bs{k}$ parallel to the surface and kinetic energy $E_\mathrm{kin}$ due to photons of energy $\hbar \omega$ and polarization vector $\bs{\lambda}$ is given by
\e{&I_{\bs{\lambda}}(\bs{k},E_{\mathrm{kin}},\hbar \omega) \lin
&\propto \sum_{i,f} |P^{if}_{\bs{\lambda}}(\bs{k})|^2  \delta(E_f - E_{i} - \hbar \omega) \ \delta(E_\mathrm{kin} - [E_f  - \phi]).
\label{eq: intensity}
}
Here, $P^{if}_{\bs{\lambda}}(\bs{k})  = \braket{f,\bs{k}|\bs{\lambda}\cdot\bs{p}|i,\bs{k}}$ is the matrix element of the momentum operator $\bs{p}$ between initial ($i$) and final ($f$) Bloch states of energy $E_i$ and $E_f$, respectively, and $\phi$ is the work function. \cite{3step}

In the simplest case of normal incidence along $\bs{e}_z$, the relevant matrix elements are $P^{if}_{\pm}(\bs{k}) = \braket{f,\bs{k}|p_x \pm i p_y|i,\bs{k}}$,~ with the signs corresponding to the two different circular polarizations of the light. In this work, we ignore final-state effects and focus on the dichroism that originates from initial states at the Fermi level, i.e., we study the Fermi-level dichroic signal defined by $D_s(\bs{k},\varepsilon_{\textrm{F}}) := I_+(\bs{k},\varepsilon_{\textrm{F}}) - I_-(\bs{k},\varepsilon_{\textrm{F}})$. To disentangle initial- and final-state information from the experimental CD spectra, one typically needs to analyze the dependence of the spectra on the photon incidence angle and on the light energy.~\cite{CDEbert1,CDEbert2,CDEbert3} In monolayer graphene unambiguous information on the initial states can be extracted from the spectra because in this system the electronic wavefunction near the Fermi level corresponds to a pseudospin that winds around each Dirac point, resulting in dichroism for a narrow range of photon energies.\cite{CDgraphene1,CDgraphene2}


We propose that the dichroism in bilayer WTe$_2$ similarly originates from \emph{tilted} Dirac fermions with small masses. As explained in \s{sec:origin}, the pseudospin rotates around the electron-like Fermi circle and in the opposite sense around the hole-like Fermi circle.
thus we expect that $D_s(\bs{k},\epsilon_{\textrm{F}})$ should carry opposite sign along the electron-like and the hole-like Fermi circles..\cite{CDgraphene1} To support this hypothesis, we calculated $D_s(\bs{k},\varepsilon_{\textrm{F}})$ using the DFT-derived initial-state wavefunctions, as further elaborated in App.\ \ref{app: CD}. As shown in \fig{fig_8}(a), $D_s(\bs{k},\varepsilon_{\textrm{F}})$ is constrained by time-reversal and $\bar{M}_x$ symmetries as $D_s(\bs{k},\varepsilon_{\textrm{F}}) = - D_s(-\bs{k},\varepsilon_{\textrm{F}})$ and  $D_s(k_x,k_y,\varepsilon_{\textrm{F}}) = - D_s(-k_x,k_y,\varepsilon_{\textrm{F}})$, respectively. When the dichroic signal is strong, $D_s(\bk,\varepsilon_{\textrm{F}})$ shows the expected sign change between electron and hole Fermi circles; we note that the signal vanishes in a small momentum segment along the electron Fermi circle. \fig{fig_8}(b) further illustrates the non-uniform variation of the $D_s(\bk,E)$ over a small range of energies.

We suggest also that the observed dichroism in 3D \wt~originates from the tilted, massive Dirac fermions. Quantitative comparison with experiment would need to account for final-state effects, spin-orbit coupling, as well as the 3D coupling between bilayers. We defer this to a future investigation. 
In closing this section, we note that ARPES and quantum oscillation experiments suggest the presence in of a zone center Fermi pocket in 3D \wt, which is absent in our DFT calculation, where it is pushed below the Fermi level. These fine effects are, however, very sensitive to small changes in the atomic parameters and functional approximations. \cite{WTe2arpes,Wte2CD,FS_calc_discussion_2}

\section{Discussion} \la{sec:discussmr}

The recent discovery of three-dimensional (3D) Dirac and Weyl semimetals\cite{liu2014stable,quinncd3as2,neupane2014observation,liu2014discovery,xu2015observation,TaAstheory1,TaAstheory2,TaAsexp1,TaAsexp2,TaAsexp3,TaAsexp4,TaAsexp5,TaAsexp6} have moved
topological semimetals to the  forefront of theoretical and experimental studies,\cite{WeylVanderbilt,WeylBalents,WeylXiaoliang,chiraltheory,chiraltheory2,xiong2015signature,NbAsexp1,NbPexp1,TaPexp1} Weyl nodes in 3D quasimomentum space require no symmetry beyond that of discrete translations, two-dimensional (2D) Dirac nodes require an additional symmorphic space-time symmetry, which in graphene is a simultaneous inversion of space and time. The robustness of Weyl and Dirac points originates from quantized topological invariants that relate to the Abelian Berry gauge field of Bloch bands.

Among the predicted Weyl semimetals, 3D MTe$_2$ (M$=$W,Mo) reveal a particularly rich range of phenomena: (i) they become superconducting under pressure,\cite{MoTe2bandgap,MoTe2Weyl1,MoTe2Weyl2,WTe2supercond1,MoTe2supercond2} (ii) \wt~demonstrates a giant, non-saturating {transverse} magnetoresistance,  and also (iii) circular dichroism in its photoemission.\cite{Wte2CD}  In this work, we demonstrate that some of these exotic 3D properties may be extrapolated from topologically characterizing a single monolayer without spin-orbit coupling (SOC); this simplification is possible because 3D \wt~has weak SOC, and is moreover composed of weakly-coupled monolayers.

We find that the spin-orbit-free MTe$_2$ monolayer belongs to a new class of band-inverted semimetals, which are diagnosed by a topological invariant associated to a {non-Abelian} Berry gauge field, which contrasts with previous Abelian Berry-type characterizations of topological semimetals. The Dirac crossings of our semimetals rely on a nonsymmorphic symmetry; they differ from previously-proposed, nonsymmorphic semimetals\cite{elementaryenergybands,Young} whose semimetallicity is guaranteed solely by the electron filling. The Dirac cones of MTe$_2$ tilt over and are classified as type-II. This has important implications for the bandstructure of 3D MTe$_2$: nearly compensated electron and hole pockets emerge, which encircle Dirac fermions with small masses; these pockets are characterized by a rotating pseudospin with consequences for circular dichroism.

We further relate our findings to the giant, non-saturating magnetoresistance observed in 3D MTe$_2$. The magnitude of the magnetoresistance in a two-band model with perfectly compensated electron and hole carriers is given by $[{\rho(H)-\rho(0)}]/{\rho(0)} = \mu_e \mu_h B^2$, where $\mu_e$, $\mu_h$ are the electron and hole mobilities.\cite{kittel2004introduction,Wte2nature}  The geometric mean of the mobilities has been extracted as $\sqrt{\mu_e \mu_h} = 167 000 \ \text{cm}^2/\text{Vs}$ in 3D WTe$_2$,\cite{EPLWTe2} as a result of the large magnetoresistance in this model. 
Additionally, the measured residual resistivity (RR) of $10^{-7} \ \Omega\text{cm}$ is extremely low compared to other binary compounds.
Applying a magnetic field of 1 T to \wt~increases the low-temperature RR to the order of $10^{-4} \ \Omega \text{cm}$.\cite{LaSbTRS,LaSbTRS2}
Extremely high mobilities are common to Dirac semimetals such as graphene, where the winding pseudospin suppresses backscattering from  impurities and defects. We suggest that a similar mechanism suppresses backscattering in 3D WTe$_2$, which hosts massive, tilted Dirac fermions. Long relaxation times in transport, in combination with the low impurity concentration in measured crystals, may account for the large mobilities. The tilting of the Dirac cone hints at the strong anisotropy in the measured MR. However, a complete explanation of the large MR should account for the coupling of the magnetic field to the electronic structure, which is known to be nontrivial for other high-MR materials such as Cd$_3$As$_2$.\cite{liang2015ultrahigh}  The nature of point defects in WTe$_2$  and their effect on the electronic structure may be probed by quasiparticle interference and studied by computational methods. 

We remark that electron-electron interactions modify the electron velocities such that they tend toward rotational isotropy at low energies. If screening is sufficiently weak, interactions may induce a type-II to type-I Lifshitz transition.\cite{nagaosatilt} The strength of screening depends on the density of states at the Fermi level, which for spin-orbit-free \wt~is not negligible.

We close this paper by remarking on the role of spatial symmetries in band topology. To date, all the proposed topological insulators that have been confirmed experimentally possess symmorphic spatial symmetry.\cite{hsieh2008,Hsieh2012,Xu2012,tanaka2012,CeX} However, a material class has recently been proposed, whose band topology relies essentially on nonsymmorphic symmetries.\cite{Hourglass}  Moreover, an analogous band topology can also be realized in photonic crystals.\cite{singlediraccone} 
Nonsymmorphic spatial symmetries have also served to classify band semimetals\cite{elementaryenergybands,Young,classific} and their Fermi-liquid analogs\cite{toplutt}, and have been used to identify topologically-ordered insulators with fractionalized excitations.\cite{nonsymmsid,nonsymmroy,watanabe}  We expect that our theory of band-inverted, nonsymmorphic semimetals, introduced here, should be broadly applicable to 3D nonsymmorphic crystals, and may be generalized to include spin-orbit coupling, as we exemplify in \fig{fig_1}(d).

\begin{acknowledgments}
The authors would like to thank M. Ali, Q. Gibson, F. Tafti, W. Hu, L. Lin, T. Berkelbach, R. Cava and D. Vanderbilt for helpful discussions as well as M. Gibertini for critical help with WANNIER90. AA was supported by the Yale Postdoctoral Prize Fellowship, and in earlier stages by NSF CAREER DMR-095242, ONR - N00014\text-11\text-1-0635, MURI\text-130\text-6082,  NSF-MRSEC DMR-0819860, Packard Foundation, Keck grant, ``ONR Majorana Fermions'' 25812\text-G0001\text-10006242\text-101, and Schmidt fund 23800\text-E2359\text-FB62. LM and RC where supported by the Department of Energy grant DE-FG02-05ER46201.
\end{acknowledgments}

\appendix

\section{Topologically classifying semimetals through the Wilson loop}\la{app:wilson}

In this Appendix, we relate the quantized $\W(k_x)$-eigenvalues (i.e., the $\pm 1$ eigenvalues which number $W_{\pm}(k_x)$) to the \emph{minimal} Dirac count ($\bar{D}_{\sma{\bk}}$), which counts the minimal number of Dirac crossings on either screw-invariant interval to the right of $l(k_x)$ (dashed lines in \fig{fig:wilsonloops}(c)). Precisely, since there are two independent, minimal Dirac counts, corresponding to either the $k_y=0$ or $k_y=\pi$ screw lines, we would relate $W_{\pm}(k_x)$ to their sum and absolute difference: 
\e{ \bar{D}_{\sma{(k_x,0)}}+\bar{D}_{\sma{(k_x,\pi)}} \eq \text{max}\{W_+(k_x),W_-(k_x)\}, \lin
\big|\bar{D}_{\sma{(k_x,0)}}-\bar{D}_{\sma{(k_x,\pi)}}\big| \eq \text{min}\{W_+(k_x),W_-(k_x)\}.\la{wilsondirac}}
This derivation is intermediated by the  relation (derived in \s{sec:bandinversion}) between the minimal Dirac count and the symmetry representations of filled states:
\e{\bar{D}_{\sma{\bk}} =\f{|N_{\sma{+,\bk}}-N_{\sma{-,\bk}}|}{2}.\la{minimalD}}
From \q{genD} and (\ref{minimalD}), we further deduce \q{wilsondirac4} and, for $a \in  \Z^{\sma{\geq}}$, 
\e{ &{D}_{\sma{(k_x,0)}}+{D}_{\sma{(k_x,\pi)}} = \text{max}\{W_+(k_x),W_-(k_x)\}+2a.\la{wilsondirac2}}
If a spatial symmetry other than screw symmetry (e.g., inversion) stabilizes Dirac crossings away from the screw lines, such crossings always exist in screw-symmetric pairs. Consequently, the total number (${D}_{\sma{l(k_x)}}$) of Dirac crossings in the cylinder bounded by $l(k_x)$ and $l(\pi)$ [red-shaded region in \fig{fig:wilsonloops}(c)] satisfies \q{wilsondirac3}.

The remainder of this Appendix aims to prove \q{wilsondirac}. Let us denote by $\pdg{g}_{\bdelta}$ a spatial transformation, which transforms real-space coordinates as $ \br \rightarrow D_g \br + \bdelta$, where $D_g$ is the orthogonal matrix representation of the point-group transformation $g$ in $\R^d$.  The nonsymmorphic space groups of interest to us contain symmetry elements for which $\bdelta$ is a rational fraction\cite{Lax} of the lattice period. Each spatial symmetry constrains the Bloch Hamiltonian as\cite{Cohomological} 
\begin{align} \label{symmonhk}
\hatgdel(\bk)\,H(\boldsymbol{k})\,\hatgdel(\bk)^{\mo} = H\big(\,D_g\boldsymbol{k}\,\big),
\end{align}
where the operator $\pdg{\hat{g}}_{\bdelta}(\bk)$ may be decomposed into a phase factor and a unitary matrix as:
\begin{align} \label{actbloch}
\pdg{\hat{g}}_{\bdelta}(\bk) \equiv e^{-i (D_g \boldsymbol{k}) \cdot \bdelta }\, U_{\sma{g\bdelta}}.
\end{align} 
The set of all operators $\{\pdg{\hat{g}}_{\bdelta}(\bk)\}$ form a representation of the space-group algebra\cite{Lax} in a basis of Bloch-wave-transformed \low orbitals;\cite{lowdin1950} in short, we call this the \low representation. The nontrivial phase factor exp$(-i D_g \bk {\cdot} \bdelta)$ in $\hatgdel(\bk)$ encodes the effect of the fractional translation, i.e., the momentum-independent matrices $\{U_{\sma{g\bdelta}}\}$ form by themselves a representation of a point group. A case in point is the screw rotation $\bcx$, which is a two-fold rotation ($g=C_{2x}$) about $\be_x$, followed by a fractional lattice translation $t(\bdelta=\be_x/2)$ parallel to the rotational axis; $\bcx^2=t(\be_x)$, an integral lattice translation. $\bcx$ is a 3D symmetry of the \wt~monolayers, which extend macroscopically in $\be_x$ and $\be_y$, and have finite, atomic-scale thickness in $\be_z$. In the 2D Brillouin zone parameterized by $\bk =(k_x,k_y)^t$, the screw rotation maps $\bk \rightarrow (k_x,-k_y)^t {\equiv} D_{2x}(k_x,k_y)^t.$ Identifying $D_g$ in \q{actbloch} with $D_{2x}$, the \low representation of $\bcx$ is
\e{ \hat{\bar{C}}_{2x}(\bk) = e^{-ik_x/2}\,U_{2x,\be_x/2}, \la{repbcx}}
which satisfies the nonsymmorphic algebra ($\bcx^2=t(\be_x)$) of a screw rotation:
\e{ \hat{\bar{C}}_{2x}(D_{2x}\bk) \,\hat{\bar{C}}_{2x}(\bk) = e^{-ik_x}.}
The momentum-independent unitary matrix $U_{2x,\be_x/2}$ forms a representation of a screwless two-fold rotation, i.e., $U_{2x,\be_x/2}^2=I$. Substituting \q{repbcx} into \q{symmonhk} yields the condition  
\e{ U_{2x,\be_x/2}\,H(k_x,k_y)\,U_{2x,\be_x/2}^{-1} =H(k_x,-k_y)}
that is, for fixed $k_x$, mathematically equivalent to the Hamiltonian of a 1D crystal with spatial-inversion symmetry represented by $U_{2x,\be_x/2}$. This identification allows us to apply a known mapping between the spatial-inversion eigenvalues (of the occupied bands) and the Wilson-loop eigenvalues (for any constant-$k_x$ momentum loop, e.g., $Y\Gamma Y$). This mapping may be found in Sec.\ IIB of Ref.\ \onlinecite{AA2014}, where the inversion eigenvalues, $\pm1$,  of filled states at the inversion-invariant momenta are now identified with the branches of screw eigenvalues ($\pm $exp$(-ik_x/2)$) of filled states at $(k_x,0)$ and $(k_x,\pi)$. For the reader's convenience, we reproduce the mapping below.

\noindent \emph{Definition:} The filled states at each ${\bk} \in  \{(k_x,0),(k_x,\pi)\}$ divide into two sets (labelled by $\xi  \in \pm  1$) according to the branch of their screw eigenvalues: $\xi\,$exp$(-ik_x/2)$. In short, we refer to $\xi$ as the screw branch, and further define the size of each set as $N_{\xi,{\bk}}$. Given the list $\{ \,N_{+,(k_x,0)},N_{-,(k_x,0)},N_{+,(k_x,\pi)},N_{-,(k_x,\pi)}\, \}$, we identify the smallest of these four integers and label it as $N_{\bar{\xi},\bar{\bk}}$, where $\bar{\xi}  \in \pm 1$ denotes the screw branch of this \emph{smallest set} of states, and $\bar{\bk} \in \{(k_x,0),(k_x,\pi)\}$ its quasi-momentum. We denote by $\bar{\bk}_{c}$ the complementary quasi-momentum. To recapitulate, we have mapped
\e{ &\{ \,N_{+,(k_x,0)},N_{-,(k_x,0)},N_{+,(k_x,\pi)},N_{-,(k_x,\pi)}\, \} \longrightarrow \lin
& \myspace\{ \,N_{\bar{\xi},\bar{\bk}},N_{-\bar{\xi},\bar{\bk}},N_{+,\bar{\bk}_{c}},N_{-,\bar{\bk}_{c}}\, \}.}  
The smallest set might be empty; in cases where the smallest set is not unique, any choice between `equally smallest' sets is valid. Let us exemplify the identification of the smallest set:

\noindent (a) Consider a $f=2$ crystal with screw branches $(++)$ at $(k_x,0)$ and $(+-)$ at $(k_x,\pi)$. The smallest set is then the empty set at $(k_x,0)$ with screw branch $-1$, i.e., $N_{\bar{\xi},\bar{\bk}}=0$ with $\bar{\xi}=-1$ and $\bar{\bk}=(k_x,0)$. 

\noindent (b) Suppose $f=4$ with screw branches $(++--)$ at $(k_x,0)$ and $(+++-)$ at $(k_x,\pi)$, then $\bar{\bk}=(k_x,\pi)$  and $\bar{\xi}=-1$.

\noindent (c) For $f=2$ with screw branches $(+-)$ at $(k_x,0)$ and $(+-)$ at $(k_x,\pi)$, we may arbitrarily pick any of the four possibilities as the smallest set. 

While the identification of $N_{\bar{\xi},\bar{\bk}}$ is simple by inspection, it is eventually worthwhile to express this quantity as
\e{N_{\bar{\xi},\bar{\bk}} \eq \f{f-\big|N_{+,\bar{\bk}}-N_{-,\bar{\bk}}\big|}{2}, \ins{with} \la{worthwile}\lin
\big|N_{+,\bar{\bk}}-N_{-,\bar{\bk}}\big| \equiv&\; \text{max}\{\big|N_{+,(k_x,0)}-N_{-,(k_x,0)}\big|, \lin
&\myspace \myspace \big|N_{+,(k_x,\pi)}-N_{-,(k_x,\pi)}\big|\}, \lin
\big|N_{+,\bar{\bk}_c}-N_{-,\bar{\bk}_c}\big| \equiv&\; \text{min}\{\big|N_{+,(k_x,0)}-N_{-,(k_x,0)}\big|,\lin
&\myspace \myspace \big|N_{+,(k_x,\pi)}-N_{-,(k_x,\pi)}\big|\} }

\noindent \emph{Mapping:} Suppose that a screw-symmetric crystal is characterized by the four quantities  $\{ \,N_{\pm\bar{\xi},\bar{\bk}},N_{\pm,\bar{\bk}_{c}}\, \}$ at a particular $k_x$. The eigenspectrum of the Wilson loop $\W[l]$, with $l$ parametrized by fixed $k_x$ and $k_y  \in  [0,2\pi)$, consists of: 

\noindent(i)  eigenvalue $-\bar{\xi}$ with multiplicity $(N_{+,\bar{\bk}_{c}} - N_{\bar{\xi},\bar{\bk}})$,

\noindent(ii) eigenvalue $+\bar{\xi}$ with multiplicity $(N_{-,\bar{\bk}_{c}} - N_{\bar{\xi},\bar{\bk}})$, 

\noindent(iii) $N_{\bar{\xi},\bar{\bk}}$ pairs of complex-conjugate eigenvalues.\\

\noindent In the above examples, the $\W$-spectrum of (a) comprises one $+{}1$ and one $-1$ eigenvalue; for (b), there are one $+1$ eigenvalue,  one $-1$ eigenvalue, and one complex-conjugate pair; (c) has one complex-conjugate pair only. For a screw-symmetric crystal with filling two and  four, we tabulate the possible  mappings, for $k_x=0$, in Table~\ref{table1d2band} and~\ref{table1d4band} respectively. 

\begin{table}[ht]
	\centering
		\begin{tabular} {|c|c|c|c|c|} \hline
						$N_{\sma{+,\bk_1}}$  & $N_{\sma{+,\bk_2}}$ & $\W(k_x)$ & $\bar{D}_{\sma{l(k_x)}}$ &  ${D}_{\sma{\bk_1}}-{D}_{\sma{\bk_2}}$ mod $2$ \\ \hline \hline
			2&2 & $[\lambda_1 \lambda_1^{\ast}  \lambda_2 \lambda_2^{\ast}]$& 0&	0\\ \hline	
			3&2 & $[\lambda_1 \lambda_1^{\ast} + -]$  & 1 & 1\\ \hline			
			3&3 & $[\lambda_1 \lambda_1^{\ast} + +]$ & 2 & 0\\ \hline
			3&1 & $[\lambda_1 \lambda_1^{\ast} - -]$  & 2& 0\\ \hline		
			4&2 & $[+ + - -]$ & 2 &0\\ \hline								
			4&3 & $[+ + + -]$ & 3 &1\\ \hline
			4&1 & $[+ - - -]$ & 3 &1\\ \hline
			4&0 & $[- - - -]$ & 4 &0\\ \hline					
			4&4 & $[+ + + +]$ & 4 &0\\ \hline			
\end{tabular}
		\caption{ Topological characterization of a crystal with filling $f=4$ and either glide or screw symmetry.  The notation used here is explained in the caption of Table\ \ref{table1d2band}.
		 \label{table1d4band}}
\end{table}

It is useful to connect this mapping to the notation $W_{\pm}(k_x)$, which was defined in \s{sec:top} as the number of robust $\pm 1$ $\W$-eigenvalues, i.e.,
\e{ W_{\pm}(k_x) =  N_{\mp \text{sign}[\bar{\xi}],\bar{\bk}_{c}} - N_{\bar{\xi},\bar{\bk}}.} 
Now we finish our proof of \q{wilsondirac}, that relates $W_{\pm}(k_x)$ to the minimal numbers of Dirac crossings along the screw lines. First, 
\e{ & 2\,\text{max}\{W_+(k_x),\,W_-(k_x)\} \lin
\eq 2 \,\text{max}\{ N_{{+,\bar{\bk}_{c}}}-N_{\bar{\xi},\bar{\bk}},\; N_{-,\bar{\bk}_{c}}-N_{\bar{\xi},\bar{\bk}} \}\lin
\eq 2 \,\text{max}\{ N_{{+,\bar{\bk}_{c}}},\; N_{-,\bar{\bk}_{c}} \} -2N_{\bar{\xi},\bar{\bk}}\lin
\eq 2 \,\text{max}\{ N_{{+,\bar{\bk}_{c}}},\; N_{-,\bar{\bk}_{c}} \} -f + \big|N_{+,\bar{\bk}}-N_{-,\bar{\bk}}\big|\lin
\eq 2 \,\text{max}\{ N_{{+,\bar{\bk}_{c}}},\; N_{-,\bar{\bk}_{c}} \} - N_{{+,\bar{\bk}_{c}}}- N_{-,\bar{\bk}_{c}} \lin
&\myspace + \big|N_{+,\bar{\bk}}-N_{-,\bar{\bk}}\big|\lin
\eq \big|N_{+,\bar{\bk}_c}-N_{-,\bar{\bk}_c}\big| + \big|N_{+,\bar{\bk}}-N_{-,\bar{\bk}}\big|\lin
\eq \big|N_{+,(k_x,0)}-N_{-,(k_x,0)}\big| + \big|N_{+,(k_x,\pi)}-N_{-,(k_x,\pi)}\big|,\la{proofdirac}}
which, upon substitution of \q{genD}  with $k_x=0$, leads immediately to the first line of \q{wilsondirac}. In the first equality of \q{proofdirac} we applied (i) and (ii) in the above mapping; in the third equality we applied \q{worthwile}; in the fourth, $f=N_{+,\bar{\bk}_c}+N_{-,\bar{\bk}_c}$. Similarly, 
\e{ & 2\,\text{min}\{W_+(k_x),\,W_-(k_x)\} \lin
\eq 2 \,\text{min}\{ N_{{+,\bar{\bk}_{c}}},\; N_{-,\bar{\bk}_{c}} \} - N_{{+,\bar{\bk}_{c}}}- N_{-,\bar{\bk}_{c}} \lin
&\myspace + \big|N_{+,\bar{\bk}}-N_{-,\bar{\bk}}\big|\lin
\eq -\big|N_{+,\bar{\bk}_c}-N_{-,\bar{\bk}_c}\big| + \big|N_{+,\bar{\bk}}-N_{-,\bar{\bk}}\big|\lin
\eq \bigg|-\big|N_{+,\bar{\bk}_c}-N_{-,\bar{\bk}_c}\big| + \big|N_{+,\bar{\bk}}-N_{-,\bar{\bk}}\big|\;\bigg|\lin
\eq \bigg|\;\big|N_{+,(k_x,0)}-N_{-,(k_x,0)}\big| - \big|N_{+,(k_x,\pi)}-N_{-,(k_x,\pi)}\big|\;\bigg|,\la{proofdirac2}}
leads to the second line of \q{wilsondirac}. 

We remark that a glide reflection, composed of a reflection $(y \rightarrow-y)$ and a half-lattice translation along $\bs{e}_x$, also maps the wavevector $(k_x,k_y)\rightarrow(k_x,-k_y)$, and is represented by a unitary matrix (squaring to identity) multiplied by a phase factor exp($-ik_x/2$), just as for a screw rotation in \q{repbcx}. Consequently, our analysis in this Section is also applicable to glide reflections if we cosmetically substitute `screw eigenvalue' with `glide eigenvalue'.

\section{Nonsymmorphic criterion for the $\Z_2$-topological phase}\la{app:qshi}

To prove our criterion for the gapped, $\Z_2$-topological phase in \s{sec:qshi}, we employ the Fu-Kane obstruction formulation\cite{fu2006} of the $\Z_2$ invariant:
\e{\nu \sim \f{i}{2\pi}\bigg[-\int_{\partial \tau_{\sma{1/2}}}\text{Tr}[\bA(\bk)] \cdot d\bk +\int_{\tau_{\sma{1/2}}} F(\bk)d^2\bk \bigg], \la{z2}}
where $\sim$ denotes equality modulo two, $\tau_{\sma{1/2}}$ denotes the half Brillouin zone illustrated in \fig{fig:qshi}, $\partial \tau_{\sma{1/2}}$ denotes the oriented boundary of $\tau_{\sma{1/2}}$ (blue line with arrows in the same Figure), Tr[$\bA$] denotes the $U(1)$ Berry connection
\e{ \text{Tr}[\bA(\boldsymbol{k})] = \sum_{i=1}^f \langle {u_{i,\boldsymbol{k}}} |{\nabla_{\boldsymbol{k}}u_{i,\boldsymbol{k}}} \rangle,} 
and $F$ denotes the $U(1)$ Berry curvature
\e{ F(\bk) = \partial_{k_x}\text{Tr}[A_y(\bk)] - \partial_{k_y}\text{Tr}[A_x(\bk)].}
In the $\Z_2$ classification, even (resp.\ odd) $\nu$ identifies a trivial (resp.\ topological) phase. Implicit in the definition of $\nu$ is our use of the time-reversal-symmetric gauge, which constrains wavefunctions related by time reversal $T$. Specifically, we divide the number of filled bands, $2f$ (counting both spin components), into $f$ pairs (labelled by an index $\alpha$), such that each pair is Kramers-degenerate at the inversion-invariant momenta. Each Kramers pair of bands is further labelled by the index $I,II$, and is constrained as
\e{\ket{u_{I,\alpha,-\bk}} = T\ket{u_{II,\alpha,\bk}},\lin
 \ket{u_{II,\alpha,-\bk}} = -T\ket{u_{I,\alpha,\bk}}, \la{trsgauge}}
for all $\alpha  \in \{1,\ldots,f\}$ and $\bk  \in \tau_{\sma{1/2}}$. In the remainder of this Appendix, our goal is to derive
\e{ \nu = \f{i}{\pi} \text{ln\,det}\,[\,\W_{\uparrow}(0)\,] \;\;\text{mod}\;2, \la{final}}
where $\W_{\uparrow}(0)$ is the  Wilson loop of the $f$ filled bands for a single spin component in the spin-orbit-free semimetal.
Equivalently, $\nu$ is odd (corresponding to the $\Z_2$-topological phase) iff det$[\,\W_{\uparrow}(0)\,]=-1$, which restates criterion (ii) in \s{sec:qshi}. Having previously proven the equivalence of criteria (i-iii) in \s{sec:qshi} our claim then follows.

\subsection{Proof of \q{final}}


The nonsymmorphic symmetry (whether screw $\bcx$ or glide $\bar{M}_y$) implies $F(k_x,k_y) = -F(k_x,-k_y)$ and thus that its integral over $\tau_{\sma{1/2}}$ in \q{z2} vanishes. What remains in Eq. \eqref{z2} is the line integral of $\text{Tr}[\bA(\bk)]$ over the oriented loop $\partial \tau_{\sma{1/2}}$. A convenient choice consistent with \q{trsgauge} is a periodic gauge $\bA(k_x,k_y)=\bA(k_x,k_y+2\pi)$. Then the line integral over $\partial \tau_{\sma{1/2}}$ reduces to
\e{ \nu \sim -\f{i}{2\pi} \left[ \int_{l(0)}\text{Tr}[\bA(\bk)] \cdot d\bk-\int_{l(\pi)}\text{Tr}[\bA(\bk)] \cdot d\bk \right],\la{reduceto}}
where $l(k_x)$ are the oriented, constant-$k_x$ loops illustrated in \fig{fig:wilsonloops}(c). Taking advantage of the time-reversal-symmetric gauge (\ref{trsgauge}), we decompose the $U(1)$ Berry connection as
\e{ \text{Tr}[\bA(\bk)] =  \bA_I(\bk) + \bA_{II}(\bk), \la{deco}}
with
\e{ \bA_{s}(\bk) = \sum_{\alpha=1}^f\langle {u_{s,\alpha,\boldsymbol{k}}} | {\nabla_{\boldsymbol{k}}u_{s,\alpha,\boldsymbol{k}}} \rangle,\; \text{for}  \;s\in\{I,II\}, }
satisfying
\e{\bA_{I}(-\bk)=\bA_{II}(\bk).\la{inverttrs}} This leads to
\e{ \nu \sim -\f{i}{\pi} \left[ \int_{l(0)}\bA_I(\bk) \cdot d\bk-\int_{l(\pi)}\bA_{I}(\bk) \cdot d\bk \right].\la{twice2}}
Since any gauge transformation that preserves \q{trsgauge} must maintain the decomposition given in \q{deco}, $U(f)$ gauge transformations within the Kramers subspace $I$ are allowed, leading to
\e{\int_{l'}\bA_I(\bk) \cdot d\bk \rightarrow \int_{l'}\bA_I(\bk) \cdot d\bk +i2\pi n,}
with $n \in \Z$. The subspace $II$ transforms dependently to maintain \q{inverttrs}, therefore $\nu$ is gauge-invariant modulo two.\cite{Fu1}If we had not imposed the time-reversal-symmetric gauge, it is well-known that any loop integral of the $U(1)$ connection [exemplified by \q{reduceto}] would only be gauge-invariant modulo one.

Suppose an interpolation (parametrized by $z \in [0,1]$) exists between semimetallic ($z=0$) and gapped ($z=1$) phases, which preserves both time-reversal and nonsymmorphic symmetries, as well as the spectral gap along both $l(0)$ and $l(\pi)$. We then introduce $\bA(\bk)\rightarrow\bA(\bk;z)$ to label the connection at a particular point in the interpolation. We would like to show that two polarization quantities, defined by
\e{ \calp(l';z) \equiv \f{i}{2\pi} \int_{l'}\text{Tr}[\bA(\bk;z)] \cdot d\bk;\;\; l' =l(0),l(\pi), }
in the time-reversal-symmetric gauge, are invariant modulo two throughout this interpolation. Having shown this, we would conclude from \q{reduceto} that $\nu$, the $\Z_2$ invariant in the fully-gapped, spin-orbit-coupled phase, obeys
\e{\nu \sim \calp(l(\pi);0)-\calp(l(0);0), \la{nueuq}} 
which we evaluate with wavefunctions of the spin-orbit-free semimetal.

\noindent \emph{Proof of invariance.} Since $l'$ is mapped onto $-l'$ by the nonsymmorphic symmetry, Eqs.~\eqref{u1berry} and (\ref{quantize}) show that $\calp(l';z)$ is independent of $z$, modulo large gauge transformations that modify $\calp$ by some additive integer. The allowed gauge transformations that preserve \q{trsgauge} cannot add an even integer to $\calp$ , as we showed earlier in this Appendix.

In the spin-orbit-free limit, we may identify the Kramers indices $(I$ and $II$) with the two spin components ($\uparrow$ and $\downarrow$), for an arbitrarily chosen spin quantization axis. Additionally, using the time-reversal-symmetric gauge we express \q{nueuq} as:
\e{\nu \sim -\f{i}{\pi} \left[\int_{l(0)} \bA_{\uparrow}(\bk) \cdot d\bk -  \int_{l(\pi)} \bA_{\uparrow}(\bk) \cdot d\bk \right],}
just as we did in \q{twice2}. Now applying the identity (\ref{u1berry}) with $\W[l(k_x)] \equiv \W(k_x)$, and further adding the subscript $\W\rightarrow \W_{\uparrow}$ to remind ourselves of the spin projection,
\e{ \nu \sim \f{i}{\pi} \bigg(\; \text{ln\,det}\,[\,\W_{\uparrow}(0)\,] -  \text{ln\,det}\,[\,\W_{\uparrow}(\pi)\,] \;\bigg). }
As described in \s{sec:bandinversion}, the oriented loop $l(\pi)$ intersects the screw-invariant (or glide-invariant) points $X$ and $M$, where, at each of $X$ and $M$, the nonsymmorphic eigenvalues of filled states always comprise $\pm i$ pairs. From the mapping of App.\ \ref{app:wilson}, we deduce that all eigenvalues of $\W_{\uparrow}(\pi)$ come in complex-conjugate pairs, i.e., det[$\W_{\uparrow}(\pi)]=+1$, leading us finally to \q{final}.

\section{Derivation of the tight-binding model of monolayer MX$_2$}
\label{app: tight-binding}

\begin{figure}[!t]
 \centering
 \includegraphics[width=0.95\columnwidth]{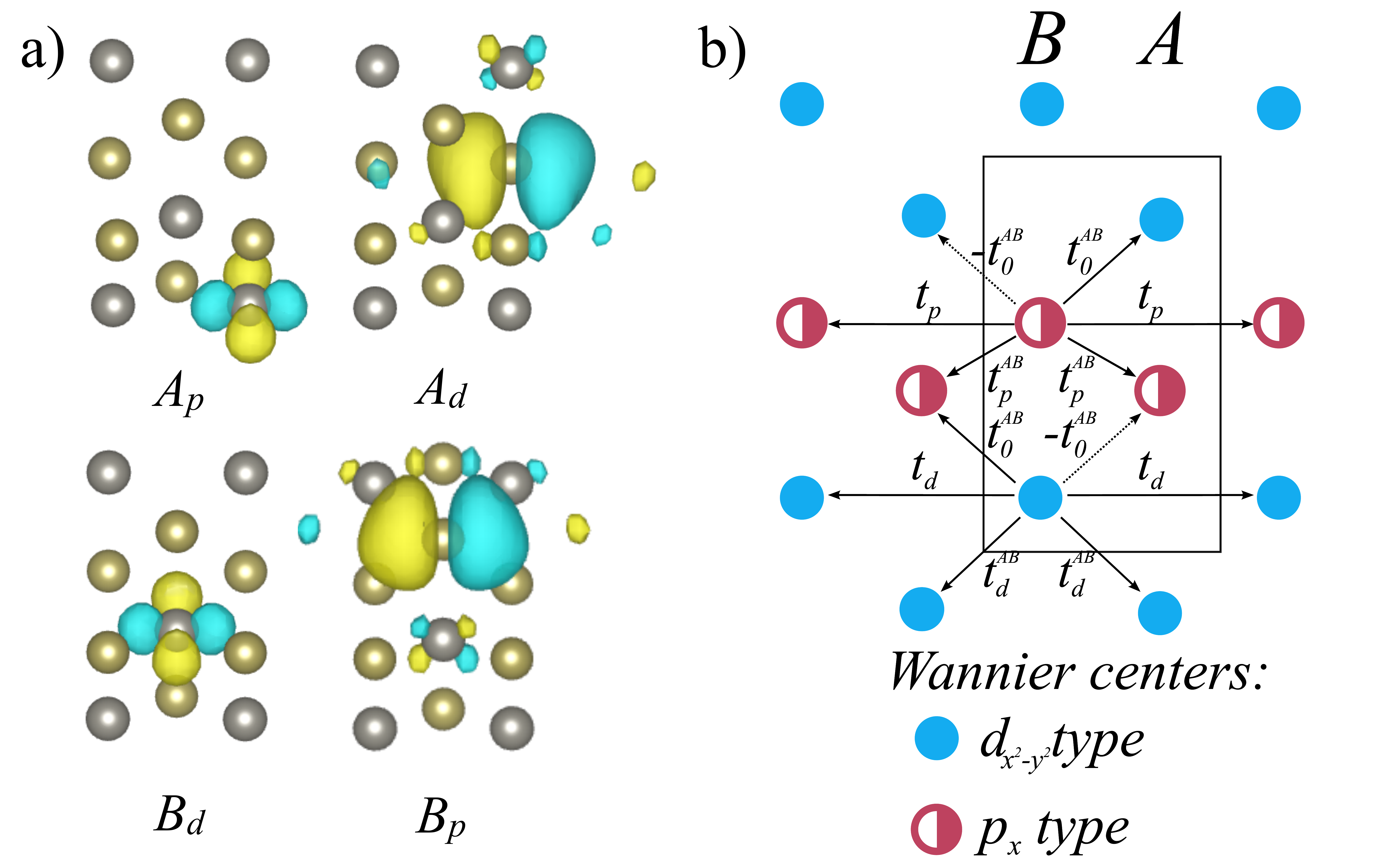}
 \caption{(a) Plots of the Wannier functions obtained by a 4-band Wannier-interpolation. The functions are labeled by their orbital character $\ell = d,p$ and their sublattice index $A,B$.
 Gray spheres represent M-atoms whereas as ochre spheres represent X-atoms. (b) Symmetry allowed hoppings considered for the tight-binding model. All hoppings are nearest-neighbor hoppings with the exception of the intra-sublattice hoppings $t_d$ and $t_p$ . All symbols below the letter A (B) belong to the A (B) sublattice.}
\label{fig_wann_appendix}
\end{figure}
\begin{table}[!t]
\parbox{.45\linewidth}{
\centering
\begin{tabular}{lrr}
\textbf{Wannier/Atom} & x/a & y/b \\
\hline
$A_d$      & -0.25  & 0.35 \\
$B_d$      & 0.25  & -0.35\\
$A_p $     & -0.25  & -0.11\\
$B_p $     & 0.25  & 0.11 \\
$W$        & 0.25  & -0.32 \\
$Te_1$        & 0.25  & 0.42 \\
$Te_2$        & 0.25  & 0.07 \\
\hline
\end{tabular}
}
\hfill
\parbox{.45\linewidth}{
\centering
\begin{tabular}{lr}
\textbf{Hopping / eV}      & \\
\hline
$\mu_d$      & 1.44   \\
$\mu_p$      & -0.38  \\
$t_d$  & -0.28  \\
$t_p$   & 0.93  \\ 
$t^{AB}_d$     & 0.52  \\
$t^{AB}_p$     & 0.40   \\
$t^{AB}_0$     & 1.02 \\
\hline 
\end{tabular}
}
\caption{\label{tab:wann_params}{Wannier functions centers and atomic positions for WTe$_2$ (left). Hopping parameters obtained from the Wannier-interpolation (right). The positions are given in units of the lattice vectors (the origin is the center of inversion), while the hopping parameters are given in eV.}}
\end{table}
To obtain a minimal, tight-binding model of the MX$_2$ compounds considered in this paper, we perform a Wannier-interpolation of the four bands closest to the Fermi level.~\cite{wannier} The Wannier functions thus obtained transform as ${d}_{x^2-y^2}$ orbitals centered close to the M atoms and ${p}_{x}$-type orbitals centered close to the X-1 atoms, as plotted in Fig.~\ref{fig_wann_appendix} and tabulated in Table ~\ref{tab:wann_params}. We remark that each Wyckoff position in the symmetry group has a fixed coordinate in $\bs{e}_x$ (corresponding to the atomic position $x/a = \pm 0.25$) but not in $\bs{e}_y$; consequently, the centers of the Wannier functions are slightly displaced in $\bs{e}_y$ from the atomic centers.

Let us consider all symmetry-allowed, nearest-neighbor hoppings, and additionally two, next-nearest-neighbor hoppings (denoted by $t_d$ and $t_p$) along the chain where atoms are closely spaced; these hoppings are illustrated in Fig.\ \ref{fig_wann_appendix}. In a basis of real Wannier functions, time-reversal symmetry constrains all hopping parameters to be real. $\bmx$ transforms the creation operators as:
\begin{equation}
\begin{split}
\bmx a^\dagger_{\ell,[R_x,R_y]} \bmx^{-1}  & = (-1)^{l} a^\dagger_{\ell,[-R_x,R_y]}, \\
\bmx b^\dagger_{\ell,[R_x,R_y]} \bmx^{-1}  & = (-1)^{l} b^\dagger_{\ell,[-R_x+a,R_y]},
\end{split}
\end{equation}
where $l = 1$ for $\ell = p$ and $l = 0$ for $\ell = d$. This suppresses nearest-neighbor hopping terms of the form $t^{AA}_{pd}a^\dagger_{p,\bs{R}} a^{\phantom{\dagger}}_{d,\bs{R}}$ and $t^{BB}_{pd}b^\dagger_{p,\bs{R}} b^{\phantom{\dagger}}_{d,\bs{R}}$. 
Given our choice of unit cell, intracell hoppings of the form $t^{AB}_{\ell,\ell'} a^\dagger_{\ell,\bs{R}} b^{\phantom{\dagger}}_{\ell',\bs{R}}$ are mapped onto $(-1)^{1-\delta_{\ell,\ell'}} t^{AB}_{\ell\ell'} a^\dagger_{\ell,[-R_x,R_y]} b^{\phantom{\dagger}}_{\ell',[-R_x+a,R_y]}$, which corresponds to an intercell hopping from the neighboring cell in the $+\bs{e}_x$ direction. For $\ell \neq \ell'$ this hopping aquires a minus sign under $\bmx$, which accounts for the factor of $(-1)^n$ in the Hamiltonian.
Spatial inversion transforms the creation operators as:
\begin{equation}
\begin{split}
\cali a^\dagger_{\ell,\bs{R}} \cali^{-1}  & = (-1)^{l} b^\dagger_{\ell,\bs{-R}} \\
\cali b^\dagger_{\ell,\bs{R}} \cali^{-1}  & = (-1)^{l} a^\dagger_{\ell,\bs{-R}},
\end{split}
\end{equation}
which enforces $t^{AB}_{\ell,\ell'} = -t^{BA}_{\ell,\ell'}$ for the interchain hoppings in the same cell with $\ell \neq \ell'$.
 
In summary, the symmetry-allowed nearest-neighbor hoppings are the intrachain hoppings $t^{AB}_{\ell,\ell} \equiv t^{AB}_{\ell} $ with $\ell = d,p$. The symmetry allowed interchain hoppings are $t^{AB}_{d,p} \equiv t^{AB}_{0}$. The latter switch sign depending on the hopping direction due to $\bmx$ and $\cali$ [see Fig.\- \ref{fig_wann_appendix}(b)]. The real-space Hamiltonian then reads:
\begin{equation} 
\begin{split}\la{tbmodapp}
H = \sum_{\bs{R}} \bigg[
&\phantom+ \sum_{\ell}  \mu_\ell \left( a^\dagger_{\ell,\bs{R}} a^{\phantom{\dagger}}_{\ell,\bs{R}}  + b^\dagger_{\ell,\bs{R}} b_{\ell,\bs{R}}^{\phantom{\dagger}}   \right) \\
&+\sum_{\ell} t_\ell \left(
 a^\dagger_{\ell,\bs{R} + \bs{e}_x } a_{\ell,\bs{R}}^{\phantom{\dagger}}  +  
b^\dagger_{\ell,\bs{R} + \bs{e}_x } b_{\ell,\bs{R}}^{\phantom{\dagger}} \right) \\
&+\sum_{n=0}^1 (-1)^nt_0^{AB} \big(  b^\dagger_{p,\bs{R} + n \bs{e}_x} a^{\phantom{\dagger}}_{d,\bs{R}}- b^\dagger_{d,\bs{R} + n \bs{e}_x} a^{\phantom{\dagger}}_{p,\bs{R}} \big) \\
&+\sum_{\ell} \sum_{\bs{\delta}_\ell} t_l^{AB} a^\dagger_{\ell,\bs{R}+\bs{\delta}_\ell} b_{\ell,\bs{R}}^{\phantom{\dagger}} 
\bigg] + \text{h.c.}.
\end{split}
\end{equation}
We construct a basis of Bloch waves by the Fourier transformation, 
\e{ &\dg{c}_{\bk,\ell,A} = \f1{\sqrt{N}} \sum_{\bR} e^{i\bk\cdot (\bR+\br_{A,\ell})} \dg{a}_{\ell,\bs{R}}, \lin
 &\dg{c}_{\bk,\ell,B} = \f1{\sqrt{N}} \sum_{\bR} e^{i\bk\cdot (\bR+\br_{B,\ell})} \dg{b}_{\ell,\bs{R}},}
where $N$ is the number of unit cells, and $\br_{s,\ell}$ denotes the position of the $\ell$-type Wannier center in sublattice $s$, as illustrated in \fig{fig:wilsonloops}(a). In the basis $[\dg{c}_{\bk,d,A},\dg{c}_{\bk,p,A},\dg{c}_{\bk,d,B},\dg{c}_{\bk,p,B}]$ the Hamiltonian is represented by the matrix:
\begin{widetext}
\begin{align}
\label{Hk}
&\mathcal{H}(\bs{k})= 
&\begin{pmatrix}
 {\mu_d}+2 t_{d} \cos({k_x})       & 0                               & t^{AB}_{d} e^{-i\bs{k}\cdot(\br_{B,d}-\br_{A,d}) } (e^{i {k_y}}+e^{i ({k_y}-{k_x})}) & t^{AB}_{0}  e^{-i\bs{k}\cdot(\br_{B,p}-\br_{A,d}) } (1-e^{-i {k_x}})\\
 0 						               & {\mu_p}+2 t_{p} \cos({k_x}) & t^{AB}_{0} e^{-i\bs{k}\cdot(\br_{B,d}-\br_{A,p}) } (e^{-i {k_x}}-1)              &  t^{AB}_{p} e^{-i\bs{k}\cdot(\br_{B,p}-\br_{A,p}) }(1 + e^{-i {k_x}}) \\
\textrm{c.c}&\textrm{c.c} & {\mu_d}+2 t_{d} \cos ({k_x})        & 0 \\
\textrm{c.c} & \textrm{c.c} &0  & {\mu_p}+2 t_{p} \cos ({k_x}) 
\end{pmatrix}.
\end{align}
\end{widetext}
\noindent 
 The relevant parameters for the Hamiltonian of WTe$_2$ obtained by the Wannier interpolation of the DFT bandstructure are given in Table~ \ref{tab:wann_params}. The Dirac crossing in the DFT bandstructure is type-II. But this is not reproduced by the tight-binding Hamiltonian \eqref{Hk} with parameters obtained from the interpolation. This is due to the truncation of longer-ranged hoppings which would further tilt the Dirac cone. To retain a minimal tight-binding model that accounts for the type-II nature of the Dirac crossing, we renormalized the intra-sublattice hoppings $t_{d},t_{p}$ while leaving the other values untouched, which results in a tilted type-II Dirac cone. Figure~\ref{fig_6}~(a) was obtained by setting $t_{d} = -0.4 $ eV and $ t_{p} = 1.34$ eV with a Fermi energy $\varepsilon_{\textrm{F}}  = 1.47 $ eV.

\section{Circular dichroism}
\label{app: CD}
To calculate the photoelectron intensity in Eq.~\eqref{eq: intensity} due to light with polarization vector $\bs{\lambda}$, one needs the transition matrix elements of the momentum operator $\bs{p}$ between the initial and final states of the electron, i.e., $P^{if}_{\pm} (\bs{k}) = \bs{\lambda}_{\pm} \cdot \braket{f,\bs{k}|\bs{p}|i,\bs{k}}$.
Within the three-step model of photoemission, $\ket{i,\bs{k}}$ and $\ket{f,\bs{k}}$ are the initial  and final Bloch state at wavevector $\bs{k}$, respectively. We assume that the polarization of the light is in the plane of the bilayer, i.e., $\bs{\lambda}_{\pm} = [1,\pm i,0]$.
The expression for the matrix element can be cast in the form 
\e{ P^{if}_{\pm}(\bs{k}) 
\eq \bs{\lambda}_{\pm} \cdot \braket{f,\bs{k}|\bs{p}|i,\bs{k}}  \lin
\eq \bs{\lambda}_{\pm} \cdot\int d\bs{r}\  e^{-i \bs{kr}} u_{f\bs{k}}(\bs{r})^{\ast} \ \bs{p}\  e^{i \bs{kr}} u_{i\bs{k}}(\bs{r}) \lin
\eq  \bs{\lambda}_{\pm} \cdot \int d\bs{r}\  e^{-i \bs{kr}} u_{f\bs{k}}(\bs{r})^{\ast} \lin 
& \qquad \times \left( \bs{k} \  e^{i \bs{kr}} u_{i\bs{k}} +  \  e^{i \bs{kr}} \ \bs{p} \ u_{i\bs{k}}(\bs{r}) \right) \lin
\eq \bs{\lambda}_{\pm} \cdot \int d\bs{r} \ u_{f\bs{k}}(\bs{r})^{\ast} \ \bs{p} \   u_{i\bs{k}}\bs{r}.} 
The Fourier expansion of the periodic part of a Bloch state is
\begin{align}
u_{n\bs{k}}(\bs{r}) = \sum_{\bs{G}} c^{[n]}_{\bs{G}\bs{k}} e^{i\bs{G}\bs{r}},
\end{align}
where $\bs{G}$ are reciprocal lattice vectors. One obtains 
\begin{align}
\la{CDformula}
P^{if}_{\pm} (\bs{k}) = \bs{\lambda}_{\pm} \cdot \sum_{\bs{G}} \bs{G} \  c_{\bs{G}\bs{k}}^{[f]\ast} c_{\bs{G}\bs{k}}^{\phantom{\ast}[i]}
\end{align}
A final state with only one nonzero coefficient $c_{\bs{G}\bs{k}}^{f}$ leads to a vanishing CD signal, because 
\begin{equation}
\begin{split}
& |P^{if}_{+}(\bs{k})|^2-|P^{if}_{-}(\bs{k})|^2  \\ 
& \qquad = \left(|G_x + i G_y|^2-|G_x - i G_y|^2 \right) |c_{\bs{G}\bs{k}}^{[f]\ast}  c_{\bs{G}\bs{k}}^{\phantom{\ast}[i]}|^2  \\
& \qquad = 0
\end{split}
\end{equation}
where $G_x, G_y$ are the components of $\bs{G}$.\cite{CDsinglePW,CDsinglePW2}
Therefore, to obtain a nonzero CD signal one  needs to use a final state with at least two nonzero plane-wave coefficients. 
It is reasonable to assume that a high energy final Bloch state has only a few nonzero plane-wave components.
Furthermore, this final state has to transform according to the symmetries of the lattice, which in case of the bilayer means that it has to transform as either an even or odd representation of $\bmx$.
In order compute the CD signal according to Eq.~\eqref{CDformula}, we chose a final state with nonzero coefficients $c_{\bs{G}\bs{k}}^{f}$ for the four smallest in-plane $\bs{G}$-vectors, i.e.,  $\bs{G} = [\pm \frac{2\pi}{a}, 0, 0],[0, \pm \frac{2\pi}{b}, 0]$.
These coefficients were assumed to be independent of $\bs{k}$ and to have the same constant value. With this assumption the final state belongs to the totally symmetric representation of $\bmx$.

The coefficients of the initial state were obtained using the plane-wave pseudopotential code in the {\sc Quantum ESPRESSO} package. For Te and W we used the pseudopotentials Te.pbe-hgh.UPF and W.pbe-hgh.UPF from the {\sc Quantum ESPRESSO} data base, respectively. The plane-wave cutoff was set to 80 Ry with a $15\times8\times1$ $k$-point mesh in the BZ.
The plane-wave coefficients were computed on a $25\times 16$ mesh around the pockets. \cite{quantumespresso}


\bibliography{WTe2}

\end{document}